\documentclass[aps,prb,twocolumn,superscriptaddress,floatfix,nobalancelastpage]{revtex4-2}
\pdfoutput=1
\usepackage[utf8]{inputenc}
\usepackage{amsmath}
\usepackage{mathtools}
\usepackage{amssymb}
\usepackage[colorlinks=true,citecolor=blue,linkcolor=blue,urlcolor=blue]{hyperref}
\usepackage[capitalise]{cleveref}
\usepackage{ulem}
\usepackage{caption}
\usepackage{subcaption}
\usepackage{times}
\usepackage[english]{babel}
\usepackage{tabularx}
\usepackage{multirow}
\usepackage{enumerate}

\usepackage{braket}

\usepackage[strict]{changepage}
\usepackage{calc}

\usepackage{graphicx}
\usepackage[export]{adjustbox}

\usepackage{xcolor}

\DeclareMathOperator{\arccot}{arccot}
\DeclareMathOperator{\diag}{diag}

\begin{document}

\title{Bulk-edge correspondence in the trimer Su-Schrieffer-Heeger model}

\author{Adamantios Anastasiadis}%
\affiliation{%
Laboratoire d'Acoustique de l'Universit\'e du Mans (LAUM), UMR 6613, Institut d'Acoustique - Graduate School (IA-GS), CNRS, Le Mans Universit\'e, France
}%

\author{Georgios Styliaris}%
\affiliation{%
Max-Planck-Institut f{\"{u}}r Quantenoptik, Hans-Kopfermann-Str. 1, 85748 Garching, Germany
}%

\affiliation{%
Munich Center for Quantum Science and Technology (MCQST), Schellingstr. 4, 80799 M{\"{u}}nchen, Germany
}%

\author{Rajesh Chaunsali}%
\affiliation{%
Department of Aerospace Engineering, Indian Institute of Science, Bangalore 560012, India
}%

\author{Georgios Theocharis}
\affiliation{%
Laboratoire d'Acoustique de l'Universit\'e du Mans (LAUM), UMR 6613, Institut d'Acoustique - Graduate School (IA-GS), CNRS, Le Mans Universit\'e, France
}%

\author{Fotios K. Diakonos}%
\affiliation{%
	Department of Physics, University of Athens, 15784 Athens, Greece}%

\begin{abstract}

A remarkable feature of the trimer Su-Schrieffer-Heeger (SSH3) model is that it supports localized edge states. Although Zak's phase remains quantized for the case of a mirror-symmetric chain, it is known that it fails to take integer values in the absence of this symmetry and thus it cannot play the role of a well-defined bulk invariant in the general case. Attempts to establish a bulk-edge correspondence have been made via Green's functions or through extensions to a synthetic dimension. Here we propose a simple alternative for SSH3, utilizing the previously introduced sublattice Zak's phase, which also remains valid in the absence of mirror symmetry and for non-commensurate chains. The defined bulk quantity takes integer values, is gauge invariant, and can be interpreted as the difference of the number of edge states between a reference and a target Hamiltonian. Our derivation further predicts the exact corrections for finite open chains, is straightforwadly generalizable, and invokes a chiral-like symmetry present in this model.

\end{abstract}
\maketitle

\section{Introduction}

SSH3~\cite{alvarez2019edge,guo2015kaleidoscope,su1981fractionally} is an extended version of the Su-Schrieffer-Heeger (SSH) model~\cite{su1979solitons,asboth2016schrieffer}. The latter has been an archetypical model for topological insulators, since it vividly displays many key aspects of these systems, while exhibiting a simple mathematical description. Specifically, it is a tight-binding model with two different sublattices within the unit cell and its hallmark is that it exhibits edge states which are robust against disorder. Moreover, a bulk-edge correspondence can be established for SSH~\cite{delplace2011zak}. Informally, this means that a topological invariant can be defined in the bulk of the system (i.e., its infinite, periodic adaption) which depends on the parameters of the Hamiltonian and takes integer values. The values of the bulk invariant correspond to the number of edge states that the finite system with open boundary conditions exhibits, while also marking its different phases.

While the SSH3 at first glance is the simplest extension of SSH, many difficulties arise concerning the investigation of the emergence of edge states and the possibility of establishing a bulk-edge correspondence. The main reason is that the 1D topological invariants which are defined in the case of SSH (e.g., Zak's phase~\cite{zak1989berry,berry1984quantal} and winding number~\cite{maffei2018topological}) require specific symmetries~\cite{rhim2017bulk} (such as chiral and inversion) in order to be quantized and to be useful in establishing an easily interpretable bulk-edge correspondence. However, these symmetries are not present in the more general case of SSH3, e.g, in the generic case when all three couplings within the unit cell are different. Moreover, in the absence of symmetries, regimes with a different total number of edge states can be adiabatically connected without necessarily being accompanied by a phase transition, in the sense of the band gap closing. However, at the same time, SSH3 is known to exhibit robust, localized edge states~\cite{alvarez2019edge}, even in the absence of symmetries, which motivates the need for an establishment of a bulk-edge correspondence. However, the difficulty in defining a bulk invariant that takes integer values for 1D systems without specific symmetries has led many to conclude that nonzero edge states are not always topological~\cite{longhi2019probing,zhu2018topological,midya2018topological}. In order to overcome the difficulties concerning defining a 1D topological invariant, a Chern number has been introduced~\cite{streda1982theory,hatsugai1993chern,thouless1982quantized} through the extension to a synthetic dimension~\cite{alvarez2019edge,zilberberg2021topology}. Other attempts for establishing a bulk-edge correspondence implement Green's function~\cite{rhim2018unified,peng2017boundary} for 1D systems.

The aim of this paper is to define a 1D bulk quantity in the infinite chain that will establish a bulk-edge correspondence. The key insight of this work is that the phases of the sublattice components of the eigenvectors of the bulk Hamiltonian contain all the necessary information for establishing a bulk-edge correspondence. Our work combines elements of the works done in  \cite{marques2020analytical,guzman2020geometry,pletyukhov2020surface,pletyukhov2020topological}. The novelty in what we present here is that we establish bulk-edge correspondence for the case of a finite open system with either integer or noninteger number of unit cells and we calculate finite-size corrections. Furthermore, the technique we use is easily generalized to larger unit cells (the case of SSH$m$). Another important aspect is that our derivation does not require any knowledge of the modern theory of polarization~\cite{spaldin2012beginner,king1993theory,resta1993macroscopic,resta1994macroscopic} which is needed in works that have used the same invariant for semi-infinite chains~\cite{pletyukhov2020surface,pletyukhov2020topological}.

Lastly, we report a new symmetry that we call \textit{point chirality}. This symmetry exists in SSH$m$ with $m$ odd and has important consequences for the behavior of the system and the profile of the edge states, similar to the ones of ordinary chirality.

\begin{figure*}[ht]
     \begin{subfigure}{0.5\textwidth}
         \centering
         \includegraphics[width=\textwidth]{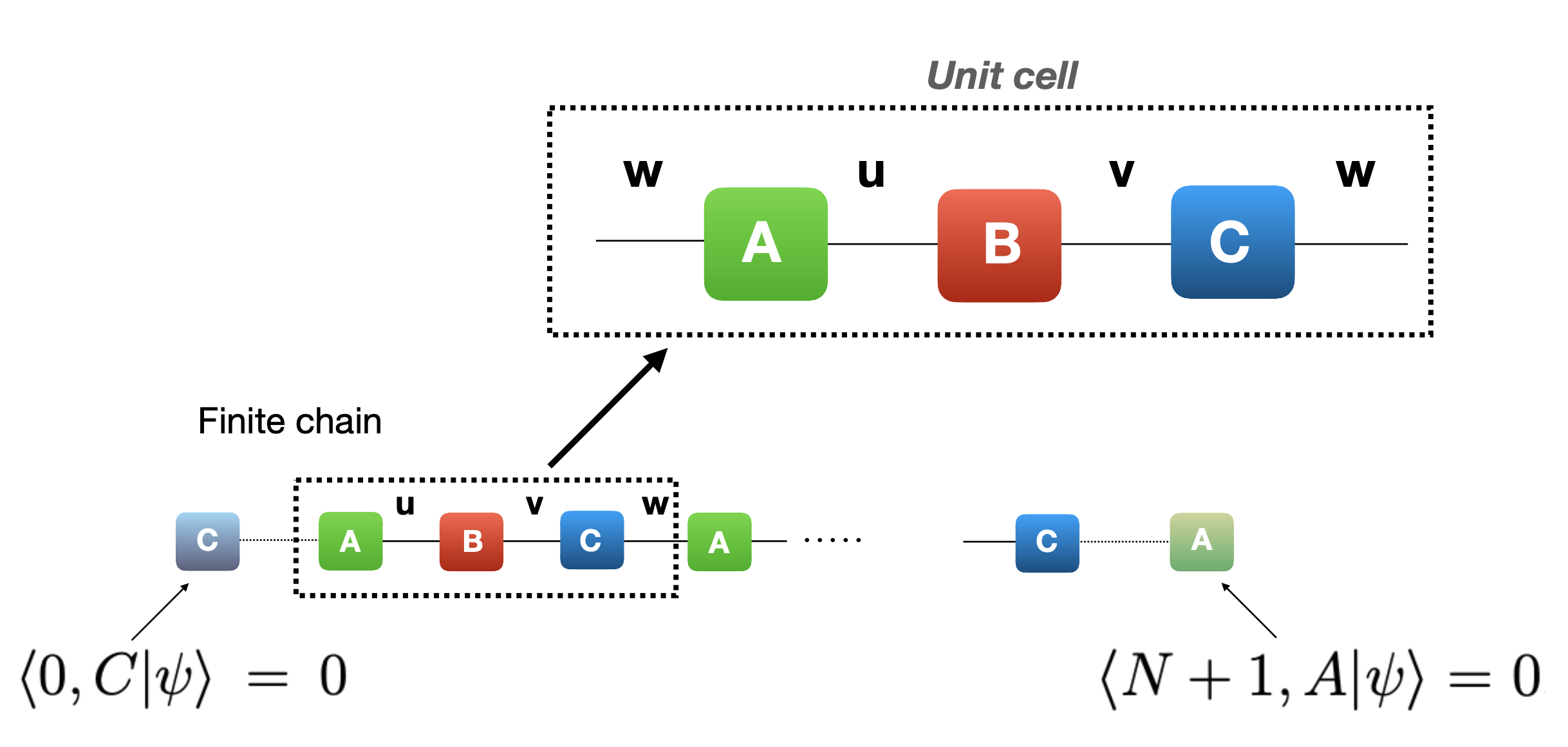}
         \caption{}
         \label{fig:periodic_fin}
     \end{subfigure}%
      \begin{subfigure}{0.45\textwidth}
         \centering
         \includegraphics[width=\textwidth]{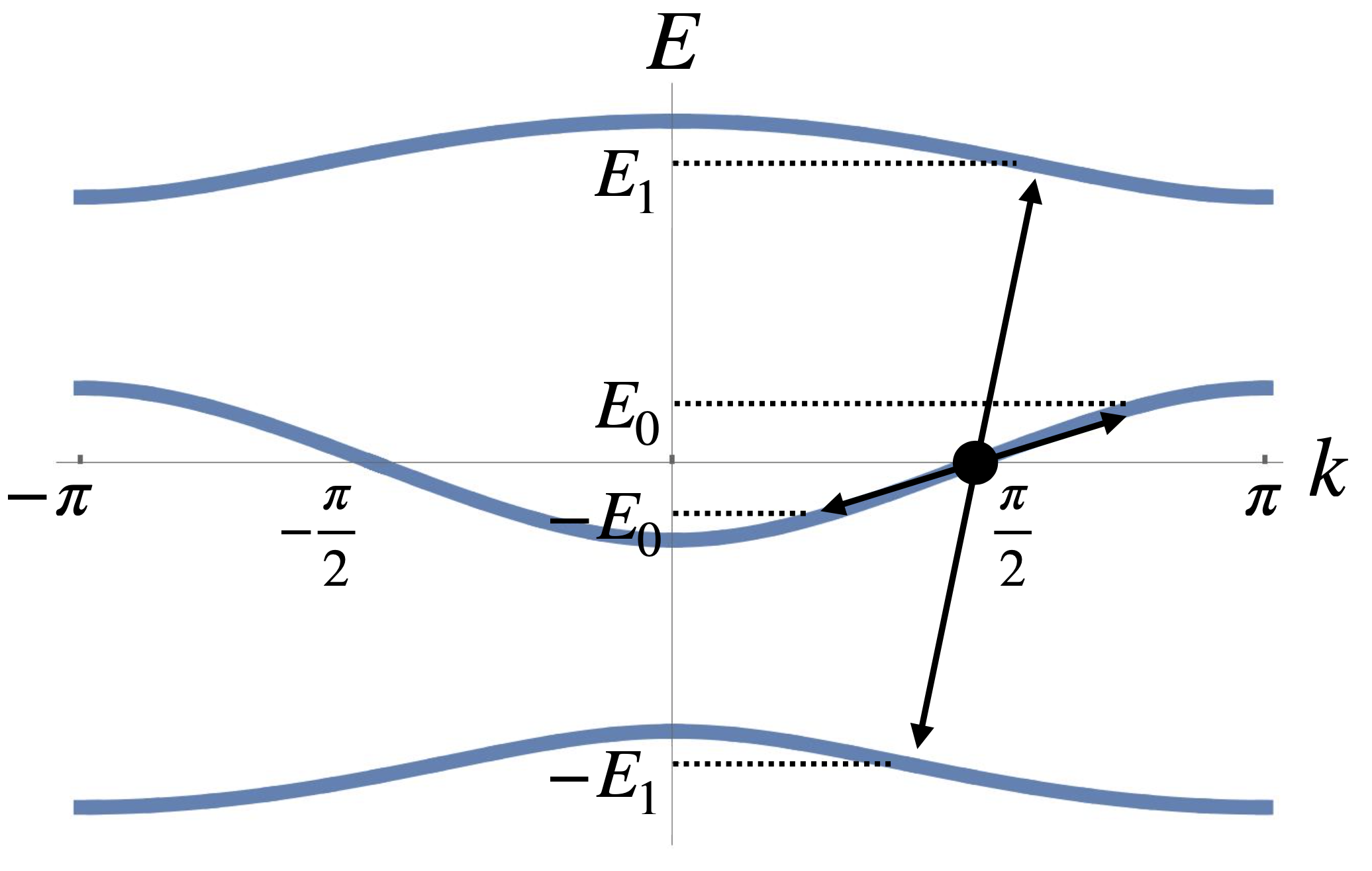}
        \caption{}
         \label{fig:spectrum}
     \end{subfigure}
            \caption{\textbf{a)}  The unit cell of an SSH3 tight-binding model with couplings $u$, $v$ (intracell) and $w$ (intercell). $A$, $B$, and $C$ denote the three sublattices while the length of the unit cell is $d$, which we set to $d=1$. 
         \textbf{b)} The spectrum of the bulk Hamiltonian of SSH3 in the first Brillouin zone. Here, all hoppings are different (no degeneracies present). The spectrum is symmetric with respect to the point $(k=\pi/2,E=0)$, as indicated.}
        \label{fig:trimer}
\end{figure*}

The rest of the paper is organized as follows: In section~\ref{sec:ssh3} we give an overview of the SSH3 model. We also present point chirality and its consequences on the spectrum and the profile of the eigenstates (including egde states). In section~\ref{sec::bulk-edge_integer} we show how the phases of the bulk eigenvectors come into play when one tries to solve the finite problem and how these phases can be obtained via normalized sublattice Zak's phase. Furthermore, we derive the proposed bulk-edge correspondence by showing that the values of normalized sublattice Zak's phase correspond to the number of edge states in each phase and we draw the phase diagram for a finite Hamiltonian with $3N$ sites. In section~\ref{sec:non_integer} we use normalized sublattice Zak's phase to establish bulk-edge correspondence for the case of chains with $3N+1$ and $3N+2$ sites. Lastly, in section~\ref{sec::Other_sshm} we apply our method to SSH and SSH4 models, hinting towards a straightforward generalization to any SSH$m$ model.

\section{The SSH3 model}\label{sec:ssh3}

\subsection{Preliminaries}

\begin{figure}[ht]
     \centering
     \begin{subfigure}[t]{0.45\textwidth}
         \centering
         \includegraphics[width=\textwidth]{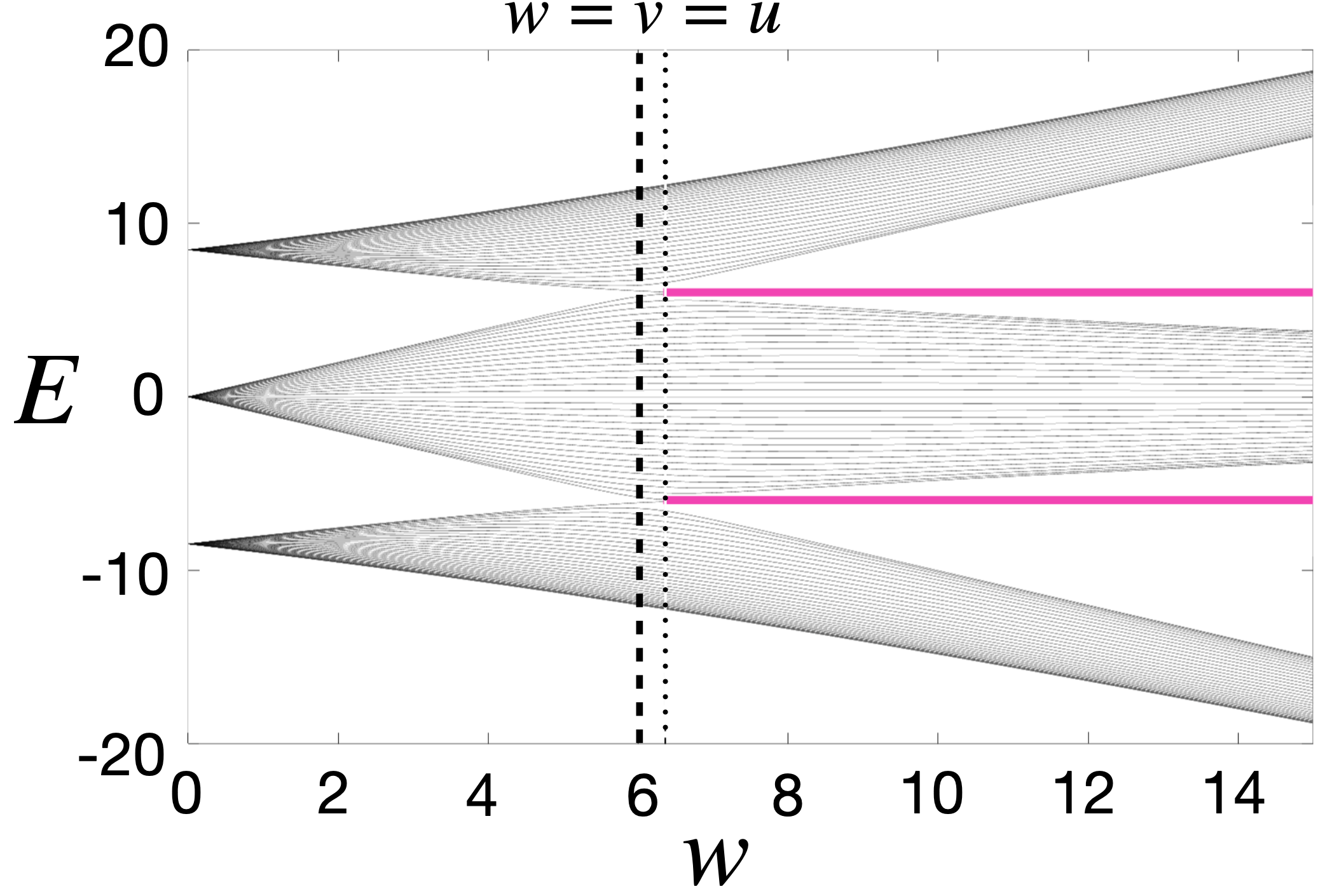}
         \caption{}
         \label{fig:yyy equals x}
     \end{subfigure}
     
     \begin{subfigure}[t]{0.45\textwidth}
         \centering
         \includegraphics[width=\textwidth]{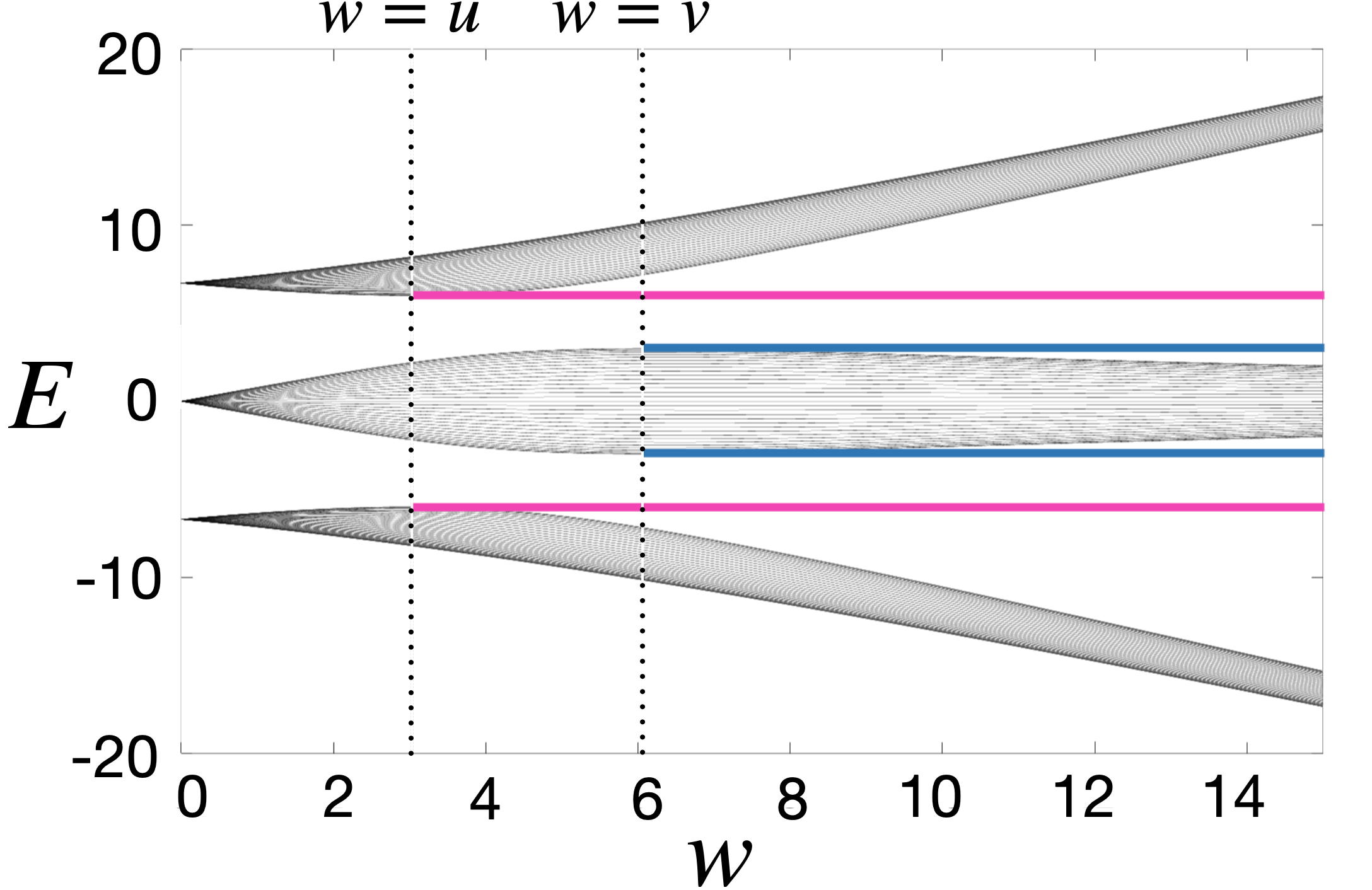}
        \caption{}
         \label{fig:three sin x}
     \end{subfigure}
            \caption{Continuation of energy spectrum with respect to the intercell hopping coupling, $w$ for a chain of $30$ sites ($n = 10$ unit cells).  \textbf{a)} A mirror-symmetric chain ($u=v=6$). Notice that the emergence of the edge states does not occur exactly at $w=6$ which would be the case in the thermodynamics limit; this is due to the finite size of the system. The exact way to calculate the finite size corrections is presented later in section~\ref{sec::bulk_edge}. \textbf{b)} A chain with all couplings different ($u = 3$ and $v = 6$) -- no mirror symmetry is present.}
        \label{fig:three graphs}
\end{figure}


SSH3 is an extended SSH model with a unit cell that consists of three sites. The hopping between the sites is controlled by the couplings $u$ and $v$, while the different unit cells are coupled with the intercell coupling $w$ (Figure \ref{fig:periodic_fin}). The system is governed by the Hamiltonian
\begin{subequations} \label{eq:hamiltonians}
\begin{multline} \label{eq:finite_open}
H = -\sum_{n=1}^{N} \big( u\ket{n,A}\bra{n,B} + v\ket{n,B}\bra{n,C} \big) \\ - \sum_{n=1}^{N-1} w\ket{n,C}\bra{n+1,A} + h.c.
\end{multline}

\noindent corresponding to an open chain, while 
\begin{multline}\label{eq:finite_periodic}
H = -\sum_{n=1}^{N} (u\ket{n,A}\bra{n,B} + v\ket{n,B}\bra{n,C}) \\ - w\ket{n,C}\bra{n \, \mathrm{mod}(N)+1,A} + h.c.
\end{multline}
\end{subequations}

\noindent for a periodic chain. Here $\ket{n,a}$ is the lattice basis, where $n = 1,2, \dots, N$ denotes the unit cell and $a = A,B,C$ denotes the sublattice. Without loss of generality, we will henceforth assume that the hopping parameters are real and nonnegative~\footnote{It is easy to verify that with a simple change of basis one can always get real and positive couplings. Assume $u = |u|e^{-i\phi_{u}}$, $v = |v|e^{-i\phi_{v}}$, $w = |w|e^{-i\phi_{w}}$. Then mapping
$\ket{m,A}\mapsto e^{-i(m-1)(\phi_{u}+\phi_{v}+\phi_{w})}\ket{m,A}$, $\ket{m,B} \mapsto e^{-i\phi_{u}}e^{-i(m-1)(\phi_{u}+\phi_{v}+\phi_{w})}\ket{m,B}$, and $\ket{m,C} \mapsto e^{-i(\phi_{u}+\phi_{v})}e^{-i(m-1)(\phi_{u}+\phi_{v}+\phi_{w})}\ket{m,C}$ one recovers the original form of the Hamiltonian with real and positive couplings.}.

The eigenstates of the periodic chain, due to translation invariance, can be constructed in terms of Bloch solutions. These solutions are of the form 
\begin{align}
    \ket{\psi_{\lambda}(k)} = \ket{k}\otimes \ket{u_{\lambda}(k)} \,,
\end{align}
\noindent where $\ket{k} = \sum_{m=1}^{N}e^{ikm}\ket{m}$ is defined over unit cells indexed by $m$ and $k$ takes values $k = 2 \pi n / N$ with $n = 1,2,\dots,N$.
The cell-periodic part $\ket{u_{\lambda}(k)}$, where $\lambda$ denotes the energy band, is an eigenvector of the reduced bulk Hamiltonian
\begin{align}\label{eq:bulk_ham}
H_{\mathrm{bulk}}(k) = - \begin{pmatrix}
0 & u & w e^{-ik}\\
u & 0 & v \\
w e^{ik} & v & 0\\
\end{pmatrix} .
\end{align}

The spectrum of the bulk Hamiltonian is not symmetric around zero, implying that the system does not possess chiral symmetry~\footnote{In this context, a system is said to have chiral symmetry (also known as sublattice symmetry) when an operator $\Gamma$ exists which is unitary and hermitian (thus $\Gamma^2 = 1$) and anticommutes with the bulk Hamiltonian, i.e, $\Gamma H(k) \Gamma^{-1} = - H(k)$. If present, chiral symmetry dictates that the spectrum is symmetric with respect to zero energy \cite{maffei2018topological}. Thus SSH3, and generally SSH$m$ models for $m$ odd, do not possess chiral symmetry.} (Figure~\ref{fig:spectrum}). Nevertheless, the bulk Hamiltonian of SSH3 possesses \textit{generalized chirality}~\cite{ni2019observation} (see Appendix~\ref{sec:app:gen_chirality_point_chirality} for a brief exposition), which relies on the fact that the system has uniform on-site potentials (i.e., the diagonal part of the Hamiltonian vanishes). 
Apart from that, the bulk Hamiltonian has an additional symmetry which we call \textit{point chirality}. 
This symmetry is a result of the equivalence of the bulk Hamiltonian of SSH3 with the bulk Hamiltonian of a degenerate SSH6 super-lattice which consists of two consecutive SSH3 unit cells \cite{he2020non}. SSH6 exhibits chiral symmetry and as a result it has a symmetric spectrum with respect to zero energy. We will show in the next section that the existence of point chirality for SSH3 gives rise to similar properties of the two systems either for the finite or for the infinite (bulk) case (i.e., for each positive eigenvalue there exists a corresponding negative eigenvalue with the same absolute value).

Furthermore, SSH3 exhibits edge states for certain values of the couplings. Edge states can appear in the case of a mirror-symmetric SSH3 (i.e., when two out of three couplings are equal) and they emerge at the point where the gap closes (Figure~\ref{fig:yyy equals x}). In general, the Hamiltonian exhibits a band gap closing only when $u=v=w$ (fully degenerate case). However, edge states can also be present in the case of an SSH3 that does not possess mirror symmetry (all couplings different) and their emergence is not related to a gap closing (Figure~\ref{fig:three sin x}). In the mirror-symmetric case it is obvious that the appearance of edge states is accompanied by the gap closing since the point where $u=v=w$ is unavoidable, while in a chain without mirror symmetry this point can be avoided. However, in the latter case, edge states appear after the passing of mirror-symmetric points, i.e., when $w = u$ and $w = v$. The aim of this work is to establish a bulk-edge correspondence that will predict the emergence of all these edge states and to explore the impact of symmetries on their profile.

\subsection{Point chiral symmetry and edge states}\label{sec:point_chiral_main}

As it is evident in Figure~\ref{fig:spectrum}, the spectrum of the Hamiltonian exhibits an interesting symmetry: For every energy $E(k)$ there exists a corresponding point at $\pi-k$ with opposite energy, possibly belonging to a different band. This observation can be formalized by noticing that~\footnote{Special care should be taken for the case of a finite periodic chain. In order for $k+\pi$ to be a valid wavenumber, the chain should have an even number of unit cells (though in the thermodynamic limit this condition becomes unimportant). If a finite periodic chain has an odd number of unit cells, (exact) point chiral symmetry breaks. However, a finite \textit{open} chain always possesses point chiral symmetry because it can also be viewed as embedded in an infinite chain upon which specific boundary conditions have been imposed (see section~\ref{sec::bulk-edge_integer}).}
\begin{align} \label{eq:point_symm}
    \Gamma_{p} H_{\mathrm{bulk}}(k) \Gamma_{p}^{\dagger} = -H_{\mathrm{bulk}}(\pi + k) 
\end{align}
\noindent where
\begin{align}
    \Gamma_{p} = \diag \left( 1, -1, 1  \right)
\end{align}
%
\noindent is unitary and hermitian. The similarity with the (ordinary) chiral symmetry is obvious, except from the fact that here the symmetry relates energy eigenstates corresponding to different $k$. This symmetry, combined with time reversal symmetry, can give a \textit{shifted} particle hole symmetry, reported in~\cite{upreti2020periodically}. An interesting consequence of this symmetry are the relations 
\begin{subequations} \label{eq:point_chiral_energy}
\begin{align}
    E_{2}(\pi - k) &= - E_{0}(k)   \\
    E_{1}(\pi - k) &= - E_{1}(k) \,,
\end{align}
\end{subequations}
\noindent where the bands $\lambda = 0,1,2$ are enumerated from bottom to top. That is, the spectrum is symmetric with respect to the \textit{point} $(k = \pi/2,E=0)$ within the Reduced Brillouin Zone (RBZ) $k\in [0,\pi]$ (analogously for $k \in [-\pi,0)$), and for this reason we will refer to symmetry~\eqref{eq:point_symm} as \textit{point chirality}. Conditions~\eqref{eq:point_chiral_energy} follow by noticing that Eq.~\eqref{eq:point_symm} is also valid for $H^*_{\mathrm{bulk}}(k)$ which, combined with time reversal, gives
\begin{align*}
    \Gamma_{p} H_{\mathrm{bulk}}(k) \Gamma_{p}^\dagger = \Gamma_{p} H^*_{\mathrm{bulk}}(-k) \Gamma_{p}^\dagger =  - H^*_{\mathrm{bulk}}(\pi -k) \,,
\end{align*}
and thus $H_{\mathrm{bulk}}(k)$ and $- H_{\mathrm{bulk}}(\pi -k)$ have the same spectrum.

For the SSH$m$ models with $m$ odd, although a chiral operator for the bulk Hamiltonian cannot be established, one can define a chiral operator for the corresponding \textit{finite open} chain as $\tilde \Gamma = \mathrm{diag}(1,-1,1,-1,1, \dots)$. In light of this observation, point chirality can be understood as a manifestation of chiral symmetry of the finite chain on the level of the bulk Hamiltonian. This becomes evident by noticing that the extension of the point chirality operator acting on the finite chain is
\begin{align}\label{eq:app:point_chiral_extension_ssh3}
    \tilde \Gamma_{p} &= \sum_{m} (-1)^{m}\ket{m}\bra{m}\otimes \begin{pmatrix}
     1 & 0 & 0\\
     0 & -1& 0\\
     0 & 0 & 1
    \end{pmatrix} \nonumber\\&= \mathrm{diag}(1,-1,1,-1,1, \dots) = \tilde \Gamma
\end{align}
\noindent where the alternating sign factor $(-1)^m$ comes from the shift $k \rightarrow k +\pi$, as prescribed by Eq.~\eqref{eq:point_symm}.

Since the symmetry~\eqref{eq:point_symm} is closely related to chiral symmetry (in the above sense), the former also implies well-established consequences~\cite{asboth2016schrieffer} for the profile of the eigenstates of the Hamiltonians~\eqref{eq:hamiltonians}. Specifically, eigenstates with nonvanishing energy satisfy the following~\footnote{As remarked earlier, for a finite periodic Hamiltonian, the number of unit cells should be even, a restriction that can be dropped for the finite periodic chain.}:
\begin{enumerate}[(i)]
    \item They always come in pairs with opposite energies,
    \item The partner eigenstates can be obtained from one another by the action of $\tilde \Gamma$,
    \item They have equal support over even and odd sites.
\end{enumerate}
For additional details and derivation of the above properties, see Appendix~\ref{sec:app:point_chiral}.

\begin{figure}[t]
\includegraphics[width=0.5\textwidth]{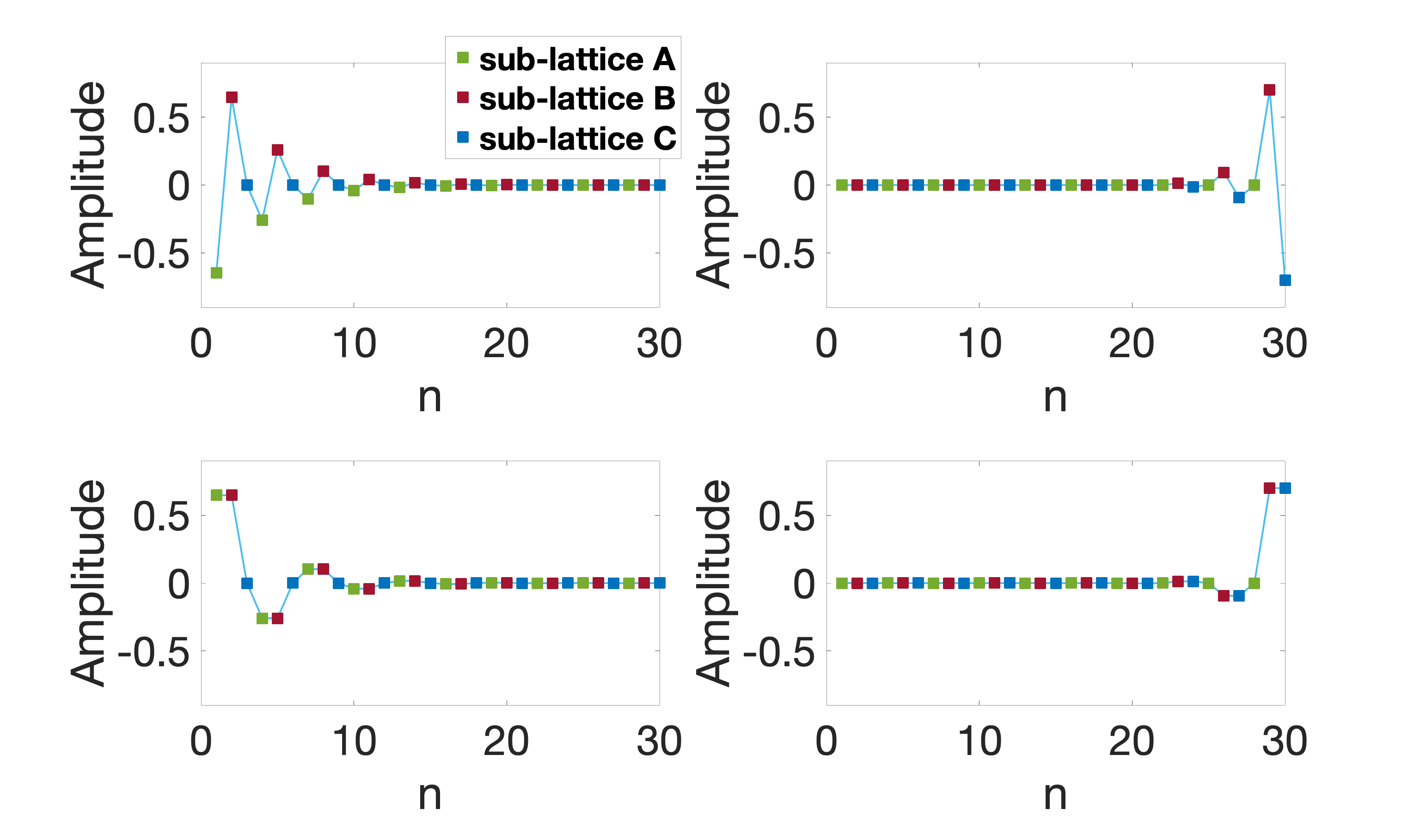}
         \caption{The spatial profile of the four edge states exhibited by an SSH3 lattice. One can observe that pairs of edge states localized on the same edge (corresponding to opposite energies) are related by a sign change of the amplitude over half of the sites. Moreover, all four edge states do not have support over one out of the three sublattices.}
         \label{fig:profile_edge}
         \end{figure}

Edge states, if they exist, can be understood as corresponding to a complex wavenumber~\cite{marques2020analytical,delplace2011zak}. Thus properties (i) - (iii) also apply to them, yielding the same features. Edge state pairs with opposite energies are localized on the same side of the chain. This follows since, by the action of a point chiral symmetry, the imaginary part of the wavenumber does not change sign. In Figure~\ref{fig:profile_edge} we plot the spatial profile of the four edge states (two in each gap) that occur for $w>u,v$ in this SSH3 chain when no mirror symmetry is present. One can observe properties (ii) and (iii) in this figure.

As mentioned previously, we aim to a bulk-edge correspondence for the SSH3 model. In practice we search for a quantized bulk quantity, which takes integer values corresponding to the number of edge states that appear in the associated open finite chain.

\section{Bulk-Edge correspondence for SSH3: Integer number of cells}\label{sec::bulk-edge_integer}

For the case of the (dimer) SSH model, the presence of inversion and chiral symmetry guarantee that Zak's phase, defined as 

\begin{align}\label{Zaks_phase}
    Z \coloneqq i\oint dk \braket{u_{\lambda}(k)|\partial_{k}u_{\lambda}(k)}
\end{align}
where the integration is carried out on the first Brillouin zone, can be a well-defined bulk quantity which takes integer values. This quantity can be used to predict the existence of edge states in each phase~\cite{delplace2011zak,asboth2016schrieffer,maffei2018topological}. The presence of these symmetries is a necessary prerequisite for the quantization of Zak's phase~\cite{guzman2020geometry,rhim2018unified}. For the case of SSH3 with $u=v$ (mirror-symmetric case) and an integer number of unit cells Zak's phase gives a correct bulk-edge correspondence \cite{alvarez2019edge}. In the absence of this constraint however, Zak's phase does not take integer values and thus cannot be directly used in order to establish a bulk-edge correspondence.

Despite the absence of the aforementioned symmetries, in the following sections we will establish a bulk-edge correspondence for SSH3 for the general case. The key ingredient will turn out to be a generalization of Zak's phase for the different sublattices, and will emerge as the natural generalization of the usual bulk invariant of the dimer SSH.

\subsection{Phases of bulk eigenvectors and normalized sublattice Zak's phase}\label{sec::finite}

We begin with the case of a finite open chain, with an integer number of cells $N$. We seek to express the eigenstates of this Hamiltonian in terms of the solutions of the periodic problem, which take the form
\begin{align}\label{bloch_big}
\ket{\psi_{\lambda}(k)} = \Lambda\sum_{j=1}^{M} e^{ikj} \ket{j}\otimes \begin{pmatrix}
a_{\lambda}^{A}(k)e^{-i\theta_{\lambda}^{A}(k)}\\
a_{\lambda}^{B}(k)e^{-i\theta_{\lambda}^{B}(k)}\\
a_{\lambda}^{C}(k)
\end{pmatrix} ,
\end{align}
\noindent where $\Lambda$ is a normalization constant to be determined after the imposition of the boundary conditions,  $M$ is the number of cells of the periodic chain and $\lambda$ enumerates the band. We assume $M \gg N $ so that $k$ can be handled as a continuous variable in $k \in [-\pi,\pi)$. Similarly to~\cite{delplace2011zak,marques2020analytical}, one considers the open chain as embedded within this longer periodic one, and imposes appropriate boundary conditions (see Figure~\ref{fig:periodic_fin}), which read
\begin{subequations} \label{eq:BC_both}
\begin{align}
    \braket{0,C|\psi_{\lambda}(k)} &= 0 \,, \label{eq:BC_1}  \\
    \braket{N+1,A|\psi_{\lambda}(k)} &= 0 \label{eq:BC_2}  \,.
\end{align}
\end{subequations}
The imposition of the boundary condition~\eqref{eq:BC_1} can be satisfied by a superposition of Bloch states in the form: $\ket{\tilde{\psi}_{\lambda}(k)} = \frac{1}{\sqrt{2}}(\ket{\psi_{\lambda}(k)} - \ket{\psi_{\lambda}(-k)})$ which can be compactly written as:

\begin{align}\label{solution}
\ket{\tilde{\psi}_{\lambda}(k)} = \tilde{\Lambda}\sum_{j=1}^{N} \ket{j} \otimes
\begin{pmatrix}
a_{\lambda}^{A}(k)\sin(kj-\theta_{\lambda}^{A}(k))\\
a_{\lambda}^{B}(k)\sin(kj-\theta_{\lambda}^{B}(k))\\
a_{\lambda}^{C}(k)\sin(kj)
\end{pmatrix} \,.
\end{align}
Here we kept only the part of the finite chain we are interested in and therefore the normalization constant $\tilde{\Lambda}$ is appropriately adapted.  To get the above form, we have also used the time-reversal symmetry, i.e, $\theta_{\lambda}^{s}(k) = - \theta_{\lambda}^{s}(-k)$ and $a_{\lambda}^{s}(k) = a_{\lambda}^{s}(-k)$ where $s = A,B,C$ and as a convenient gauge choice we fix the $C$-sublattice component to be real ($\theta_\lambda^C(k) = 0$). 

Now, by imposing the second boundary condition~\eqref{eq:BC_2}, one gets

\begin{align}\label{theta}
\theta_{\lambda}^{A}(k) = (N+1)k - n_{\lambda}\pi, \qquad n_{\lambda} \in \mathbb Z \,
\end{align}

\noindent where $n_{\lambda}$ counts the allowed real $k$ solutions. The finite problem will have exactly $3N$ eigenvectors, i.e.,  $N$ for each band. If condition \eqref{theta} yields less than $N$ distinct eigenstates for each band, the missing solutions can be expressed in terms of complex wavenumbers and turn out to be the edge states~\cite{delplace2011zak,alvarez2019edge}.

In turn, since the number of allowed solutions is dictated by \eqref{theta}, it is natural to assume that for the case of SSH3 with an integer number of unit cells, the condition that determines the number of Bloch solutions -- and thus, also the number of localized solutions -- for each band is encoded in a bulk quantity, namely the phase $\theta_\lambda^{A}$ of the first sublattice. The above reasoning dictates that this information can be extracted through the winding of a sublattice. This winding was defined in \cite{pletyukhov2020surface} and we will call it \textit{normalized sublattice Zak's phase} (NS Zak's phase) because it can be also viewed as a quantity similar to the one defined in~\cite{guzman2020geometry}, but with an extra normalization \footnote{The formula of NS Zak's phase given in \eqref{sublattice_Zak_one} can be evaluated by treating the variable $\theta_\lambda^{A}(k)$ over the real line instead of choosing a branch of range $2\pi$. This is because, if a specific branch is chosen, then Eq.~\eqref{sublattice_Zak_one} should be altered and contribution $\pi$ must be added for each jump that happens when the angle changes branch. On the other hand, $H_{\mathrm {bulk}}(k)$ has a smooth dependence on the wavenumber $k$, which is inherited to its eigenvectors (away from the degenerate point $u=v=w$). As such, the variables $\{\theta_\lambda^s(k)\}_{s=A,B,C}$ always admit a unique (smooth) extension over the real line, up to an additive constant which drops out of Eq.~\eqref{sublattice_Zak_one}. We will henceforth adopt this convention, and hence evaluate $Z_{A,C}^\lambda = \theta_{\lambda}^A (\pi) - \theta_{\lambda}^A (0)$.}:

\begin{align}\label{sublattice_Zak_one}
Z_{A,C}^\lambda & \coloneqq  \frac{i}{2}\oint dk \braket{\tilde{u}_{\lambda}(k)|\partial_{k}\tilde{u}_{\lambda}(k)} =\int_{0}^{\pi}dk \frac{\partial \theta_{\lambda}^{A}(k)}{dk} \,.
\end{align}

\noindent Above we defined $\displaystyle \ket{\tilde{u}_{\lambda}(k)} \coloneqq \frac{P_{A}\ket{u_{\lambda}(k)}}{\sqrt{\braket{u_{\lambda}(k)| P_A |u_{\lambda}(k)}}} = e^{-i\theta^{A}_{\lambda}(k)}\ket{A}$ and $P_{A} \coloneqq \ket{A}\!\bra{A}$ is the sublattice projector. The first subscript denotes the projected sublattice, the second subscript denotes the gauge condition $\theta^C_\lambda = 0$, while $\lambda$ denotes the corresponding band. We also used the fact that $\theta_{\lambda}^{A}(k)$ is an odd function of $k$ due to time-reversal symmetry in order to integrate in the RBZ.

\subsection{Normalized sublattice Zak's phase: $\mathbb{Z}$ or $\mathbb{Z}_{2}$ invariant?}

Zak's phase, as defined in Eq.~\eqref{Zaks_phase}, is interpreted by many authors as a $\mathbb{Z}_{2}$ invariant \cite{maffei2018topological,longhi2019probing,midya2018topological} especially in the context of super-lattices and many bands. This means that Zak's phase can take only two integer values that correspond to two phases (trivial or topological). In this framework, Zak's phase is defined $\mathrm{mod} (2\pi)$, the quantization is achieved via chiral or inversion symmetry \cite{zak1989berry} and it is used to characterize \textit{band gaps} as trivial or topological (if they contain edge states or not) in the following manner: In a multi-band system, one calculates the sum of Zak's phases corresponding to all the bands below the gap to be probed. The number of edge states in the desired gap is the outcome of the summation $\mathrm{mod} (2\pi)$. On the contrary, here we aim to probe bands instead of gaps. This means that instead of the question ``how many edge states are in a given gap?", we are going to answer the question ``how many states are missing from a given band?". It is then clear that the bulk quantity should be defined as a $\mathbb{Z}$ invariant (as is the case with Eq.~\eqref{sublattice_Zak_one}), i.e., an invariant that can take all integer values, and not as a $\mathbb{Z}_{2}$ invariant, because some bands can contribute two edge states (see Figure~\ref{fig:three graphs}).

In order to achieve gauge invariance, we will employ differences of NS Zak's phases. Apart from being mathematically helpful \cite{cooper2019topological}, this way of defining the invariant also corresponds to an experimentally measurable quantity \cite{atala2013direct}. Specifically, we will assign physical meaning only to a difference of NS Zak's phases with respect to an (arbitrary, but conveniently chosen) reference Hamiltonian. The NS Zak's phase for this Hamiltonian will be denoted as $Z_{A,C}^{\lambda,\mathrm{ref}}$ and it will be defined in a specific gauge. As we will momentarily show, the gauge-invariant quantity $(Z^{\lambda}_{A,C} - Z_{A,C}^{\lambda, \mathrm{ref}})/\pi$ will be equal to difference in the number of edge states (in band $\lambda$) between the target Hamiltonian and the reference one.

\subsection{Bulk-edge correspondence for a chain with an integer number of unit cells} \label{sec::bulk_edge}

We are now ready to state the main result of this section: \textit{An (open) SSH3 chain with $3N$ sites, in the thermodynamic limit $N \to \infty$, has exactly $ (Z_{A,C}^\lambda - Z_{A,C}^{\lambda,\mathrm{ref}})  / \pi$ edge states which have emerged from band $\lambda$. $Z_{A,C}^{\lambda,\mathrm{ref}}$ denotes the NS Zak's phase of the ``reference'' coupling regime that has no edge states, which here can be any chain with $w<u,v$.}

Although we will henceforth work for convenience with the specific gauge prescribed earlier, the above statement is independent of this choice. In fact, our derivation will yield a stronger result, allowing us to predict analytically the phase diagram as a function of the hopping parameters for \textit{any finite size}.

\subsubsection*{Derivation of the bulk-edge correspondence for a chain with an integer number of unit cells}

As discussed previously, the number of the delocalized eigenstates of the Hamiltonian~\eqref{eq:finite_open} is dictated by the quantization condition~\eqref{theta}. We now turn to investigate analytically the dependence of the number of distinct solutions of this equation as a function of the hopping parameters of the Hamiltonian.

We will follow the approach of Ref.~\cite{banchi2013spectral}, that is, parametrize the eigenvalues of the finite problem in terms of the eigenvalues of the bulk Hamiltonian. This can be achieved via recursive relations for the characteristic polynomials involving subdeterminants of the finite Hamiltonian. By implementing this technique, one arrives at a complete set of conditions that, when satisfied, yield the non-edge state solutions of the finite problem. A derivation for SSH$m$ ($m \in \mathbb N$) can be found in~\cite{marques2020analytical}. The specific condition for the case of SSH3 is derived for convenience in Appendix~\ref{sec:app:momentum_shift} and reads:
\begin{subequations} \label{eq:subeqs:central}
\begin{gather}\label{central}
\cot(\phi_{\lambda}(k)) = \cot[(N+1)k] = \frac{1}{a^{1}_{\lambda}(k)\sin(k)} + \cot(k)  \\
a^{1}_{\lambda}(k) \coloneqq - \frac{w}{uv}E_{\lambda}(k) \,,
\end{gather}

\noindent where the angular variable $\phi_{\lambda}(k)$ is known as the \textit{momentum shift}~\cite{marques2020analytical} and $E_{\lambda}(k)$ is the energy eigenvalue of the band $\lambda$. For the case of an integer number of cells and gauge $\theta^{C}_{\lambda}(k) = 0$, one can in addition take

\begin{align}\label{momentum_shift_sublattice}
\phi_{\lambda}(k) = \theta^{A}_{\lambda}(k) \,.
\end{align}
\end{subequations}

We are now able to deduce (i) the value of the NS Zak's phase and (ii) the number of edge states over the hopping parameter space.

\subsubsection*{Normalized sublattice Zak's phase}

By combining Eqs.~\eqref{sublattice_Zak_one} and \eqref{eq:subeqs:central} it immediately follows that \textit{NS Zak's phase is quantized}, i.e., $Z_{A,C}^\lambda = n_\lambda \pi$  with $n_\lambda \in \mathbb Z$. This is because
\begin{align}\label{central_2}
\cot[\theta^{A}_{\lambda}(k)] =  \frac{1 - \frac{w}{uv}E_{\lambda}(k) \cos (k)}{-\frac{w}{uv}E_{\lambda}(k) \sin(k)}
\end{align}

\noindent and, since the eigenenergies $E_{\lambda} (k)$ are continuous and bounded functions of $k$, Eq.~\eqref{central_2} diverges when $k \to 0^+, \pi^-$.

Our next task is to deduce the values $Z_{A,C}^\lambda$ as a function of the Hamiltonian parameters. For that, we will first determine the hypersurfaces in the parameter space that separate regions with different values of $Z_{A,C}^\lambda$. In Appendix~\ref{sec:app:trigonometric} we show that all such separating hypersurfaces are determined by the equation $g_{\lambda}(k)=0$, where $g_{\lambda}(k) \coloneqq 1 - \frac{w}{u v} E_{\lambda}(k) \cos(k)$ is the numerator in Eq.~\eqref{central_2}.

The values of the hopping parameters satisfying the equation $g_{\lambda}(k)=0$ can be obtained analytically via invoking the fact that the eigenvalues of the bulk Hamiltonian satisfy the characteristic polynomial. The latter reads

\begin{align}\label{char_pol_bulk}
E_{\lambda}^{3} (k) - (u^{2}+v^{2}+w^{2}) E_{\lambda} (k) + 2 u v w \cos(k) = 0 \,.
\end{align}

\noindent With this equation at hand, the condition $g_{\lambda}(k)=0$ then immediately reduces to
\begin{subequations}
\begin{gather}\label{parameters}
a(a+1)(a-1)b(b+1)(b-1) = 0, \quad \text{where} \\
a \coloneqq \frac{u}{w}, \quad \text{and} \quad b \coloneqq \frac{v}{w} \,,
\end{gather}
\end{subequations}

\noindent independently of the band $\lambda$.

We have therefore shown that all possible changes of the quantized $Z_{A,C}^\lambda$ can occur on the lines $a=0,\pm1$ and $b = 0, \pm 1$ of the $a - b$ real plane, i.e., when $w = \pm u$ and $w = \pm v$. The parameter space is divided by the above surfaces into regions over which NS Zak's phase, for all of the bands, cannot change value. It is hence now straightforward to deduce the actual value of $Z_{A,C}^\lambda$  for each band, and for each of the few resulting regions of the parameter space. This can now be done easily (see Appendix~\ref{sec:app:trigonometric}), and the resulting values for NS Zak's phase are summarized in Table~\ref{tab:table}.

\begin{center}
\begin{table}[t]
\begin{tabular}[t]{ |c||c|c|c|  }
 \hline
 &$Z_{A,C}^\lambda - Z_{A,C}^{\lambda,\mathrm{ref}}$ & $Z_{A,C}^\lambda - Z_{A,C}^{\lambda,\mathrm{ref}}$&$Z_{A,C}^\lambda - Z_{A,C}^{\lambda,\mathrm{ref}}$ \\
 Band $\lambda$ & $a,b >1 $ &  $a<1 $, $b>1$ & $a,b<1$ \\
 & & $a>1$, $b<1$& \\
 \hline 
  $0$   & 0    &$\pi$&   $\pi$\\
$1$&   0 & 0  & 2$\pi$\\
$2$ & 0 & $\pi$ &  $\pi$\\
 \hline
\end{tabular}
\caption{Table of the different values of $Z_{A,C}^\lambda - Z_{A,C}^{\lambda,\mathrm{ref}}$ for the case of integer number of unit cells in different coupling regimes. Here the reference chain is any chain with $a,b>1$.} \label{tab:table}
\end{table}
\end{center}

\begin{figure*}[t]
     \centering
     \begin{subfigure}[t]{0.45\textwidth}
         \centering
         \includegraphics[width=\textwidth]{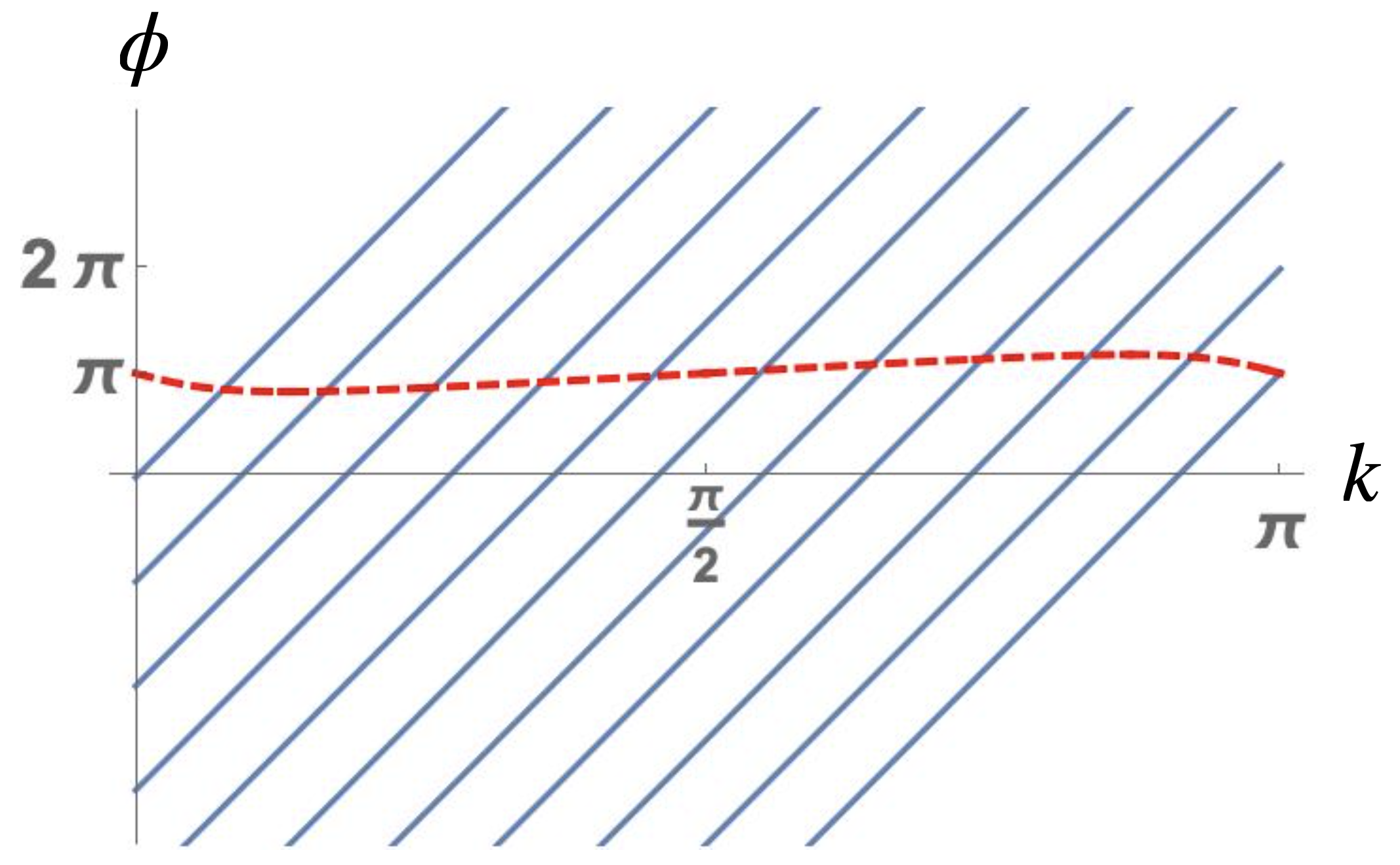}
        \caption{}
         \label{fig:y equals x}
     \end{subfigure}
     ~
     \begin{subfigure}[t]{0.45\textwidth}
         \centering
         \includegraphics[width=\textwidth]{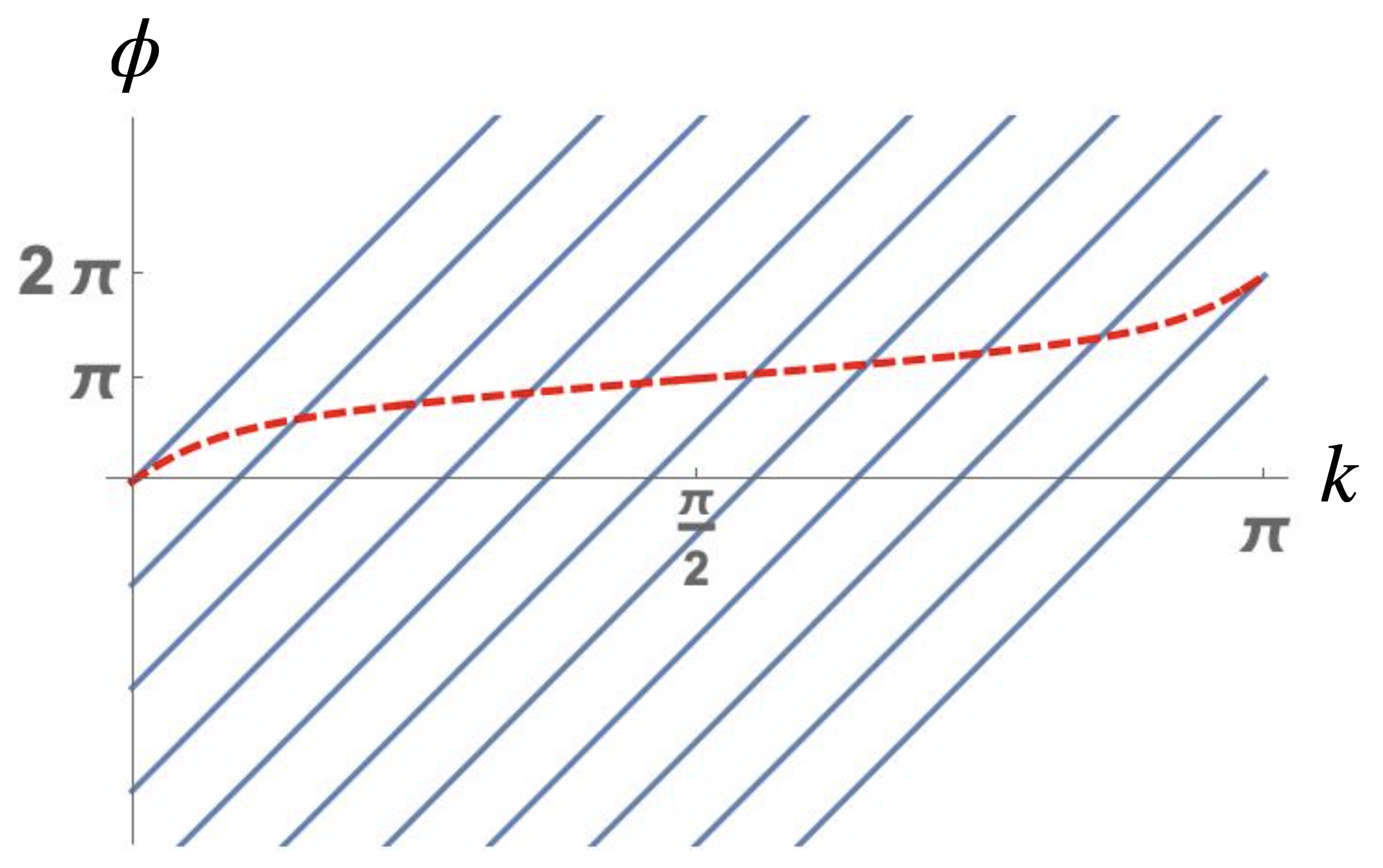}
       \caption{}
         \label{fig:analytical_phi}
     \end{subfigure}
     
     \begin{subfigure}[t]{0.5\textwidth}
         \centering
         \includegraphics[width=\textwidth]{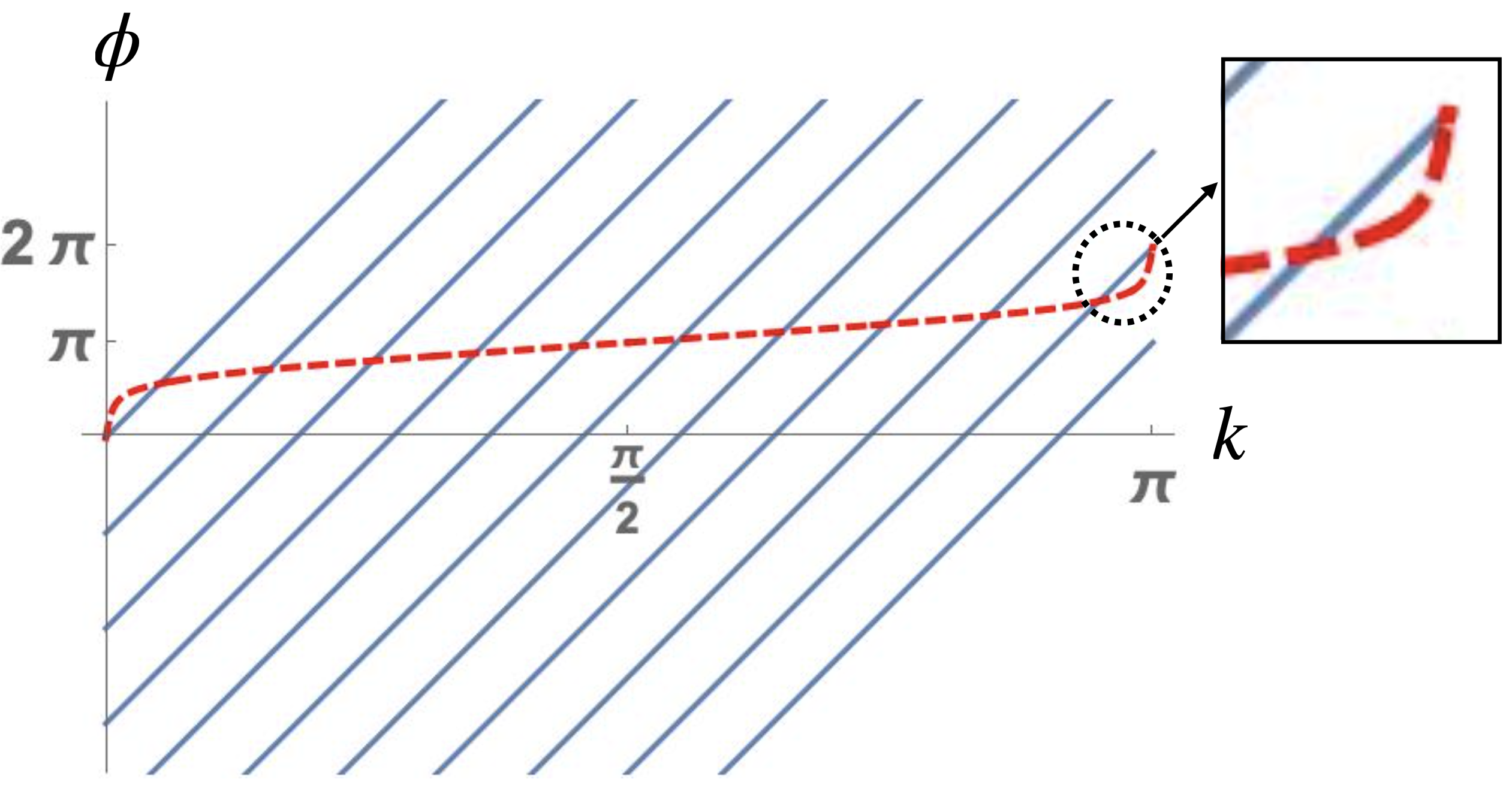}
        \caption{}
         \label{fig:error}
     \end{subfigure}
            \caption{Plot of the momentum shift $\phi_{\lambda}(k)$ (red dashed line) from its analytical expression (see Appendix~\ref{sec:app:trigonometric}) for the middle band in two different coupling regimes for a finite chain with $N=10$ unit cells. \textbf{a)} In the trivial regime, one can observe that this bulk quantity intersects with the $\{F_{n}(k)\}$ lines (blue) 10 times in the Reduced Brillouin zone $k\in(0,\pi)$. \textbf{b)} When in the edge states regime, momentum shift intersects 8 times with the possible solutions of the finite problem. Two states have left the band. \textbf{c)} Finite size effect, where NS Zak's phase is nonzero but the same number of intersections occurs as in the trivial regime. This kind of finite size effect vanishes in the thermodynamic limit ($N \rightarrow \infty$).
            }
        \label{fig:phi}
\end{figure*}

\subsubsection*{Counting edge states}

The last piece remaining is to show how one can systematically count the number of edge states as a function of the hopping parameters, also taking into account finite-size effects. For that, we will follow an approach similar to Ref.~\cite{delplace2011zak}.

Effectively, this task reduces to counting the number of distinct eigenstates yielded by $\theta_{\lambda}(k)$ when Eq.~\eqref{theta} is taken into account. Graphically, one can equivalently investigate the number of intersections of $\phi_{\lambda}(k)$ with the lines $\{ F_{n}(k) \}_n$, where $F_{n}(k) \coloneqq (N+1)k - n \pi$, $n \in \mathbb Z$ in the \textit{open interval} $k\in (0,\pi)$. The reason that one should take the RBZ is that due to time-reversal symmetry, only half of the Brillouin zone produces distinct eigenstates. Furthermore, one has to exclude the endpoints of the interval because the eigenstates are identically zero at $0$ and $\pi$ again due to time-reversal symmetry. Therefore, in the trivial regime -- where the system does not exhibit edge states -- the number of intersections between $\phi_{\lambda}(k)$ and the lines $\{ F_{n}(k) \}_n$ in the RBZ should be exactly $N$, i.e., equal to the number of unit cells of the finite system we are investigating. On the other hand, if the system exhibits edge states, this means that there are less than $N$ intersections within the RBZ.

This is demonstrated in Figure~\ref{fig:phi} for the middle band ($\lambda = 1$). Due to the continuity of $\theta_\lambda^A (k)$, each band cannot contribute with more than two edge states, i.e., missing solutions always correspond to endpoint lines. Moreover, the number of edge states (for an arbitrary finite size $N$) does not only depend on the values at the endpoints, but also on the shape of the curve near the endpoints. In fact, the relevant feature is the slope (see Figure~\ref{fig:phi}), and the corresponding condition is analogous to the one stated by Delplace et al.~\cite{delplace2011zak}. Specifically, the slope condition takes the form
\begin{subequations}
\begin{align}\label{condition}
0\le \partial_{k}\theta^A_{\lambda}(k)|_{k=\pi} \le \partial_{k}F_{n}(k)|_{k=\pi}
\end{align}
\noindent which reduces to
\begin{align} \label{eq:slope_condition_simplified}
   1\geq \frac{1}{a_{\lambda}^{1}(\pi)} \geq 1 - \frac{1}{N+1} \,
\end{align}
\end{subequations}

\noindent where $\lambda = 1,2$  denotes the middle or the top band, respectively. 
The above condition is satisfied if and only if a) there is a double edge state contribution from the middle band (case $\lambda = 1$) and b) there is a single edge state contribution from the top band and a single one from the bottom bands (case $\lambda = 2$).

It is important to note that due to point chiral symmetry, the behavior of the bottom band can be deduced from the top band, hence there is no need to examine the case $\lambda = 0$ separately.

Point chiral symmetry imposes the following constraints on the momentum shift: a) The momentum shift of the middle band is symmetric around $(k, E) = (\pi/2,0)$. Thus, the fact that the middle band \textit{necessarily} contributes with pairs of edge states can be understood as a consequence of the above unitary symmetry.  b) The momentum shift of the bottom band is symmetric to the one of the topmost band with respect to $(k, E) = (\pi/2,0)$. As a result, when the condition for the existence of an edge state is satisfied for either the top or the bottom band, a corresponding relation will be satisfied for the other one as well.

It is easy to see that in the thermodynamic limit $N\rightarrow\infty$, $\partial_{k}F_{n}(k)|_{k=\pi}\rightarrow\infty$. This means that condition \eqref{eq:slope_condition_simplified} will be always satisfied as long as momentum shift is a smooth function with a positive finite value at $k=\pi$. In that case, only the values of the momentum shift at the edges of the RBZ are relevant and the derivative conditions can be dropped. 
As a result, the values of  $Z_{A,C}^\lambda$ are sufficient to determine alone the number of edge states in the thermodynamic limit.

\begin{figure}
\includegraphics[width=0.45\textwidth]{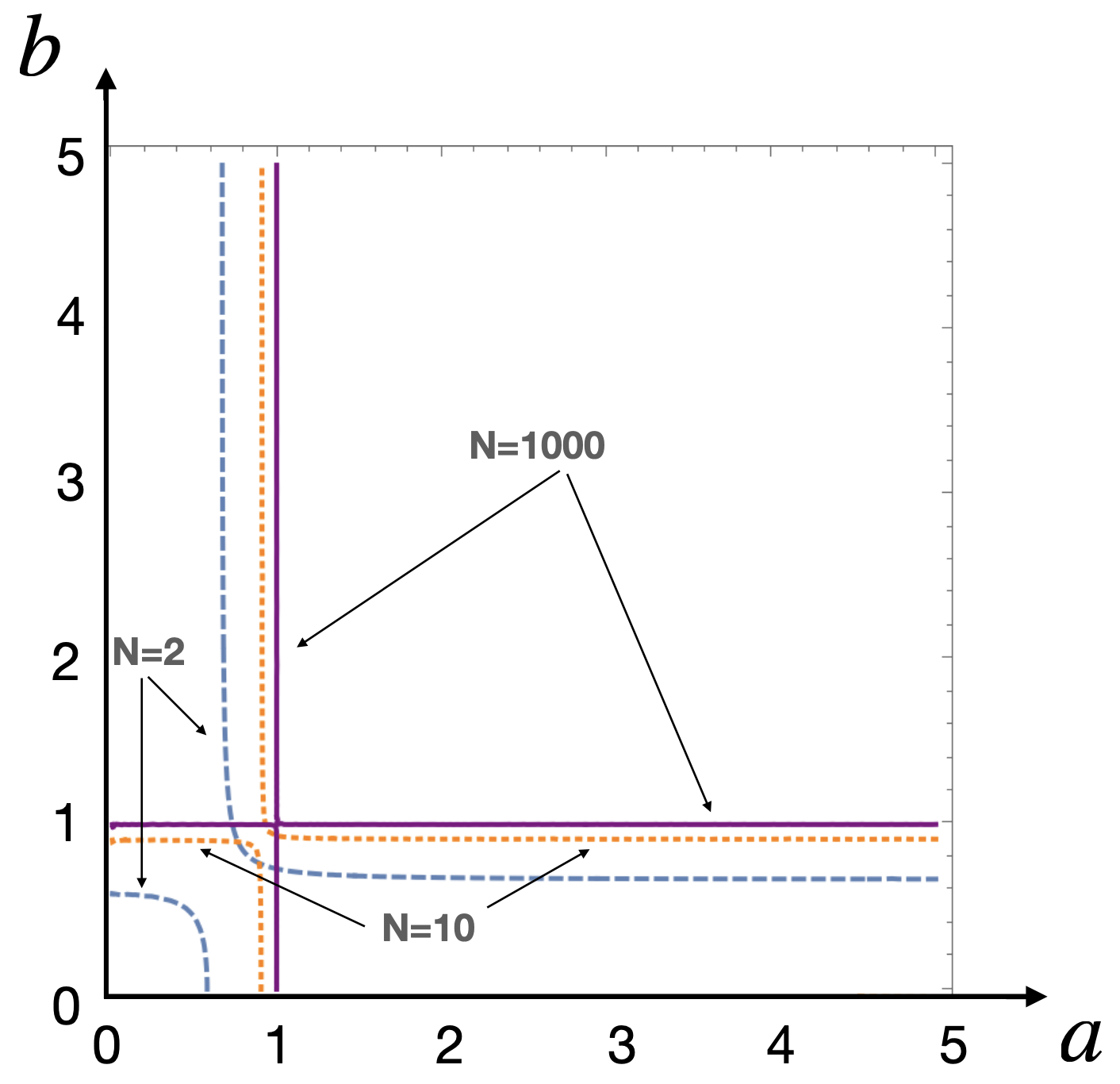}
         \caption{Plot of the limiting lines from relation~\eqref{limit_lines} where the transition happens for the finite system, in the parametric space of $a$ and $b$, for the bottom and middle band. Chains have $N=2,10,1000$ (dashed line, dotted line, and solid line respectively) unit cells. Notice how, at the thermodynamic limit, the curves approach the lines predicted by \eqref{parameters}.  Comparing the current figure with Fig.~\ref{fig:phase_diagram} it becomes evident that the number of edge states matches with the values predicted analytically by NS Zak's phase.} 
         \label{fig:contours}
         \end{figure}

Now, coming back to \eqref{eq:slope_condition_simplified}, one can investigate the limiting case where:
\begin{align}\label{limit_lines}
    \frac{1}{a_{\lambda}^{1}(\pi)} =  1 - \frac{1}{N+1} \,.
\end{align}
This defines size-dependent curves in the $(a,b)$-parameter space separating regions where a different number of edge states are exhibited, see Figure~\ref{fig:contours}. In the thermodynamic limit, one recovers the phase diagram predicted by the NS Zak's phase, since the above equation reduces to $g_{\lambda}(\pi) = 0$, concluding the derivation of the bulk-edge correspondence for an integer number of unit cells.

\begin{figure}
\includegraphics[width=0.5\textwidth]{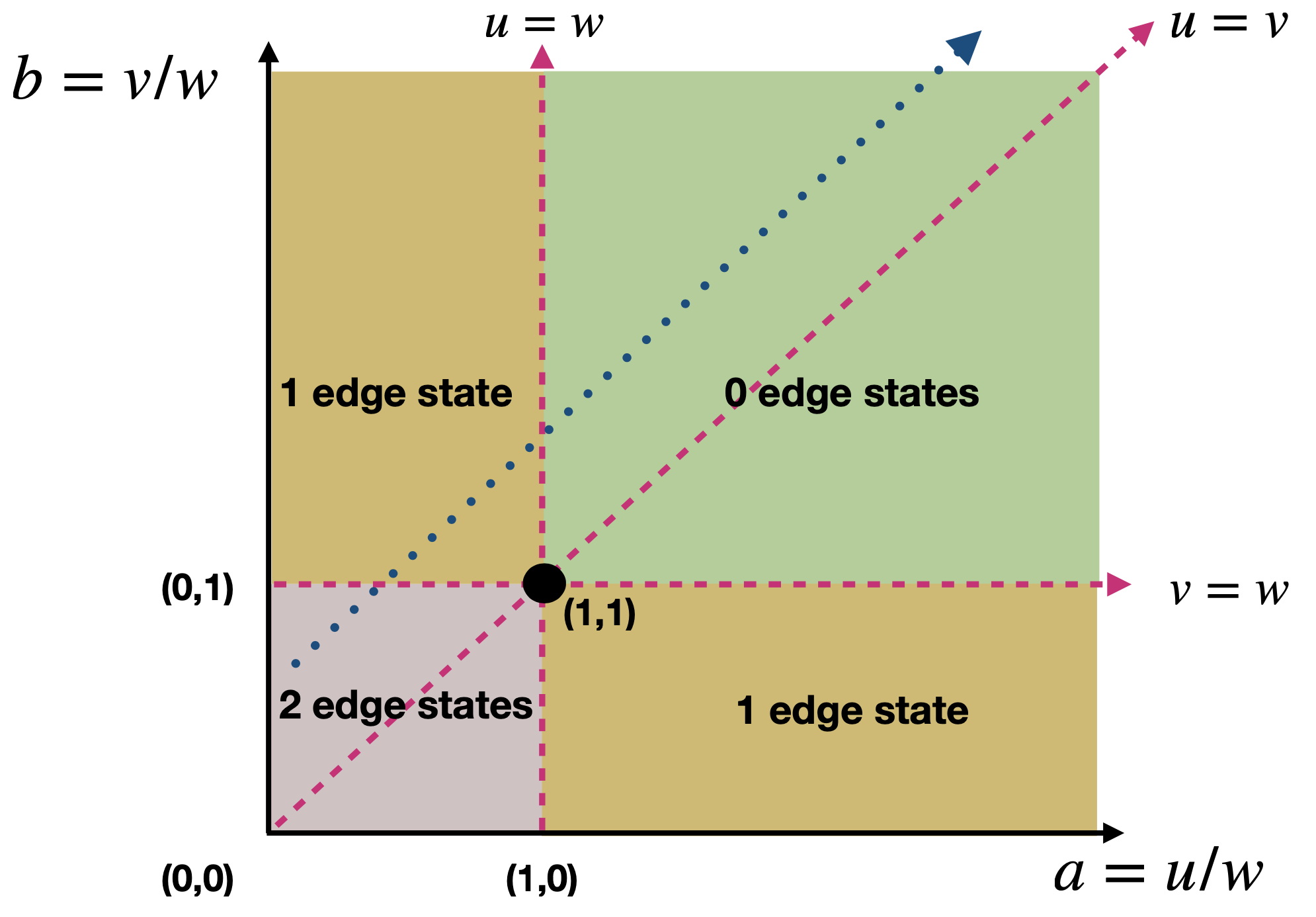}
         \caption{Parameter regimes with a different number of edge states for an SSH3 chain with $3N$ sites, as predicted by the introduced bulk quantity. Dashed lines represent the mirror-symmetric paths while the path with the dotted lines is not constrained by mirror symmetry.}
         \label{fig:phase_diagram}
         \end{figure}

In Figure~\ref{fig:phase_diagram} the parameter regimes with  different number of edge states as predicted by the introduced bulk quantity are presented. Only the number of edge states below zero energy (first band gap) are shown. The number of edge states above zero energy will be the same due to point chiral symmetry. As one can observe, the closing of the gap happens only at the single point $(a=1,b=1)$. Thus, it is not necessary for the path of an adiabatic change of the Hamiltonian to pass through $(1,1)$ when the number of edge states changes. The agreement of the size-dependent curves of Figure~\ref{fig:contours} at the thermodynamic limit with the borderlines in Figure~\ref{fig:phase_diagram} verifies the bulk-edge correspondence introduced in the present work.

\section{Bulk-edge correspondence for $3N+1$ and $3N+2$ sites}\label{sec:non_integer}

The key observation that allowed us to establish a bulk-edge correspondence in the case of $3N$ sites was that the phase of the first component of the Bloch eigenvectors (in the appropriate gauge) codified all the necessary information for the existence of the edge states. However, when extra sites are added the expression for the momentum shift is modified~\cite{marques2020analytical}, as dictated by the corresponding boundary conditions. It is then natural to pursue a formulation of a bulk-boundary correspondence that is given in analogous terms, i.e., as a NS Zak's phase but over a different sublattice (see Figure~\ref{fig:different_bc}). One would also expect that a gauge-invariant quantity can be constructed by taking the difference of the NS Zak's phases between the chain of interest and a reference chain, as in the case of $3N$ sites.
Specifically we will show that all the differences  $ (Z_{i,C}^\lambda - Z_{i,C}^{\lambda,\mathrm{ref}})  / \pi$ with $i = A,B,C$ can be utilized to establish bulk-edge correspondence for different case of non-commensurate chains. 

In the following we show that, indeed, a similar procedure leads to a correct bulk-edge correspondence for the case of $3N+1$ sites. With the appropriate modifications, the derivation is exactly analogous to the case of $3N$ sites. On the other hand, the case of $3N+2$ sites needs individual treatment, a fact that has also been observed in more general context~\cite{marques2020analytical,he2018topology}.

\subsection{Bulk-edge correspondence for a chain with $3N+1$ sites}

When adding an extra site to the right of a chain with an integer number of cells, the boundary conditions~\eqref{eq:BC_both} are modified as (see Figure~\ref{fig:different_bc})
\begin{subequations} \label{eq:BC_both_non_integer}
\begin{align}
    \braket{0,C|\psi_{\lambda}(k)} &= 0 \,, \label{eq:BC_1_non_integer}  \\
    \braket{N+1,B|\psi_{\lambda}(k)} &= 0 \label{eq:BC_2_non_integer}  \,.
\end{align}
\end{subequations}
By following exactly the same logic as in the integer case, we arrive at the quantization condition:
\begin{align}\label{theta_b}
\theta_{\lambda}^{B}(k) = (N+1)k - n\pi, \qquad n \in \mathbb Z \,.
\end{align}
This implies that the proper bulk quantity in this case is $Z_{B,C}^\lambda$ and not $Z_{A,C}^\lambda$, as this NS Zak's phase extracts the relevant angle in the case of $3N+1$ sites.
If one chooses the gauge
\begin{align}\label{new_gauge}
\theta_{\lambda}^{C}(k) = 0 \,,
\end{align}
then the additional relation
\begin{align}
  \theta_{\lambda}^{B}(k) =   \phi_{\lambda}(k)
\end{align}
holds, where $ \phi_{\lambda}(k)$ is the momentum shift. The latter, in the case of an extra site, takes slightly different form:
\begin{subequations} \label{eq:subeqs:central_two}
\begin{gather}\label{central_new}
\cot(\phi_{\lambda}(k)) = \cot[(N+1)k] = \frac{1}{a^{2}_{\lambda}(k)\sin(k)} + \cot(k)  \\
a^{2}_{\lambda}(k) \coloneqq - \frac{uw}{v}\frac{1}{E_{\lambda}(k)} \,.
\end{gather}
\end{subequations}
\noindent For a thorough investigation of the momentum shift for non-integer unit cells, see \cite{marques2020analytical}.

Similarly to the approach in \ref{sec::bulk_edge}, one arrives at the hypersurfaces on which NS Zak's phase changes value: $a'=\pm1$, $b'=\pm1$ in the $a'-b'$ space, where now $a' \coloneqq \frac{u}{v}$ and $b' \coloneqq \frac{w}{v}$.

\begin{figure}[t]
\includegraphics[width=0.5\textwidth]{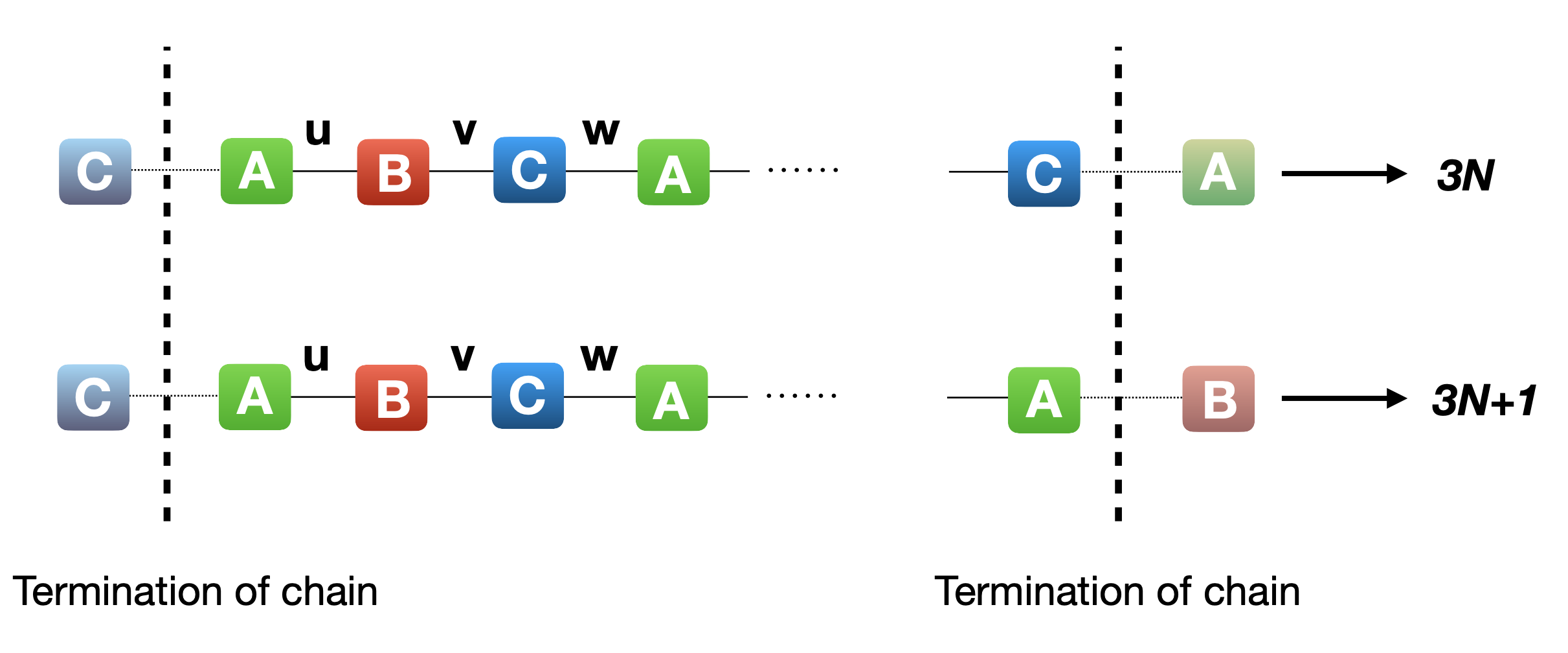}
         \caption{The boundary conditions of an open chain for the cases of $3N$ and $3N+1$ sites. Since they are not the same, a different NS Zak's phase should be utilized in each case.}
         \label{fig:different_bc}
         \end{figure}

For the case of a finite chain with $3N+1$ sites, one can use the same technique for counting states within the Brillouin zone as in the case of the integer number of unit cells. The analytical formula for the momentum shift is given in Appendix~\ref{sec:app:trigonometric}, from which one can calculate NS Zak's phase for the different parameter regimes. 

Lastly, in order to get finite size corrections one should again use the derivative condition given in \eqref{condition} expressed for $\theta^B_{\lambda}$. This reduces to
\begin{align} \label{eq:slope_condition_simplified_ni}
   1\geq \frac{1}{a_{\lambda}^{2}(\pi)} \geq 1 - \frac{1}{N+1} \,
\end{align}
which is the exact analogue of \eqref{eq:slope_condition_simplified}. In the light of these, we arrive at the bulk-edge correspondence for an SSH3 chain with ${3N+1}$ sites:

\textit{An (open) SSH3 chain with $3N+1$ sites, in the thermodynamic limit $N \to \infty$, has exactly  $ (Z_{B,C}^\lambda - Z_{B,C}^{\lambda,\mathrm{ref}})  / \pi$ edge states corresponding to the band $\lambda$. $Z_{B,C}^{\lambda,\mathrm{ref}}$ denotes the NS Zak's phase of the ``reference'' coupling regime that has no edge states, which here can be any chain with $v>u,w$.}

\begin{center}
\begin{table}[t]
\begin{tabular}[t]{ |c||c|c|c|  }
 \hline
 & $Z_{B,C}^\lambda - Z_{B,C}^{\lambda,\mathrm{ref}}$ & $Z_{B,C}^\lambda - Z_{B,C}^{\lambda,\mathrm{ref}}$&$Z_{B,C}^\lambda - Z_{B,C}^{\lambda,\mathrm{ref}}$ \\
 
 Band $\lambda$  & $a',b' >1 $ &  $a'<1 $, $b'>1$ & $a',b'<1$\\
 &&$a'>1$, $b'<1$& \\
 \hline
$0$ & $\pi$    & 0 &   $0$\\
$1$ &   $2\pi$ & $2\pi$  & $0$\\
$2$ &$\pi$ & 0 &  0 \\
 
 \hline
\end{tabular}
\caption{Difference of NS Zak's phases $Z_{B,C}^\lambda - Z_{B,C}^{\lambda,\mathrm{ref}}$ for $3N+1$ sites in different regimes. The reference chain now is a chain with $a',b'<1$. It is worth noting that $Z_{B,C}^{\lambda,\mathrm{ref}}$ is not always zero, but the difference gives the correct number of edge states (see Figure \ref{fig:triankaiena}). } \label{tab:table_two}
\end{table}
\end{center}

\begin{figure}[t]
\includegraphics[width=0.5\textwidth]{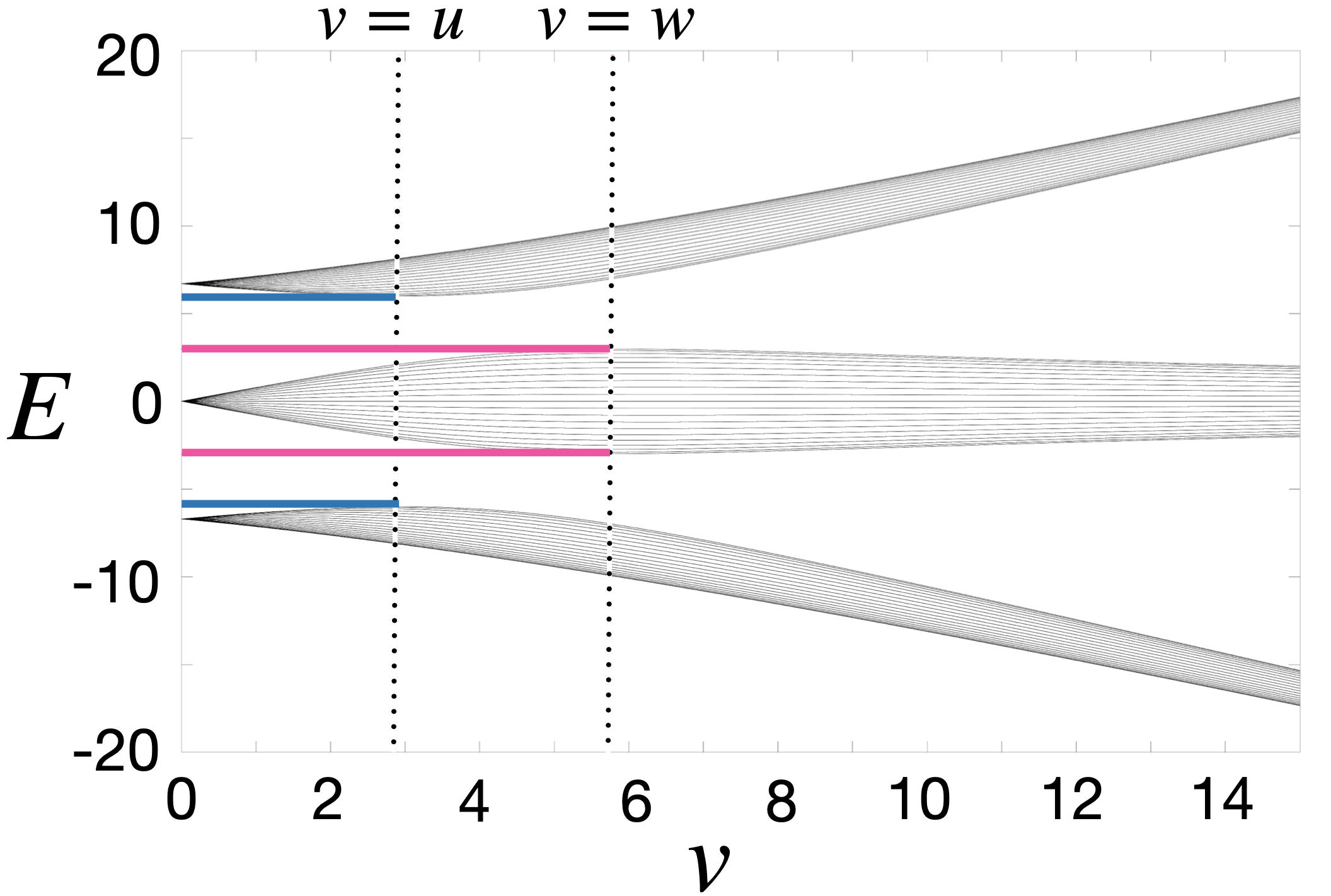}
         \caption{Continuation diagram for a finite SSH3 chain with 3N+1 sites.}
         \label{fig:triankaiena}
         \end{figure}

Notice that, in the case of $3N$ sites, the hypersurfaces where the borderlines between parameter regions with different number of edge states were given by $a,b = \pm1, 0$ with $a = \frac{u}{w}$ and $b = \frac{v}{w}$. What this means is that by changing adiabatically $w$, one could probe all the transitions (emergence and disappearance of edge states) in a continuation diagram of the spectrum with respect to $w$. However, in the case of $3N+1$ sites, the hypersurfaces of the transitions are given by $a',b' = \pm1$ with $a' = \frac{u}{v}$ and $b' = \frac{w}{v}$. This means that a continuation diagram with respect to $w$ will miss a transition. Thus the continuation diagram that captures all the transitions is with respect to $v$ in this case, as shown in Figure~\ref{fig:triankaiena}. This diagram verifies the predictions of NSZ for the case of $3N+1$ which are presented in Table~\ref{tab:table_two}. By taking this into account, it is easy to create the phase diagram for the case of $3N+1$ sites on the $a'-b'$ plane.

\begin{figure}[t]
         \includegraphics[width=0.5\textwidth]{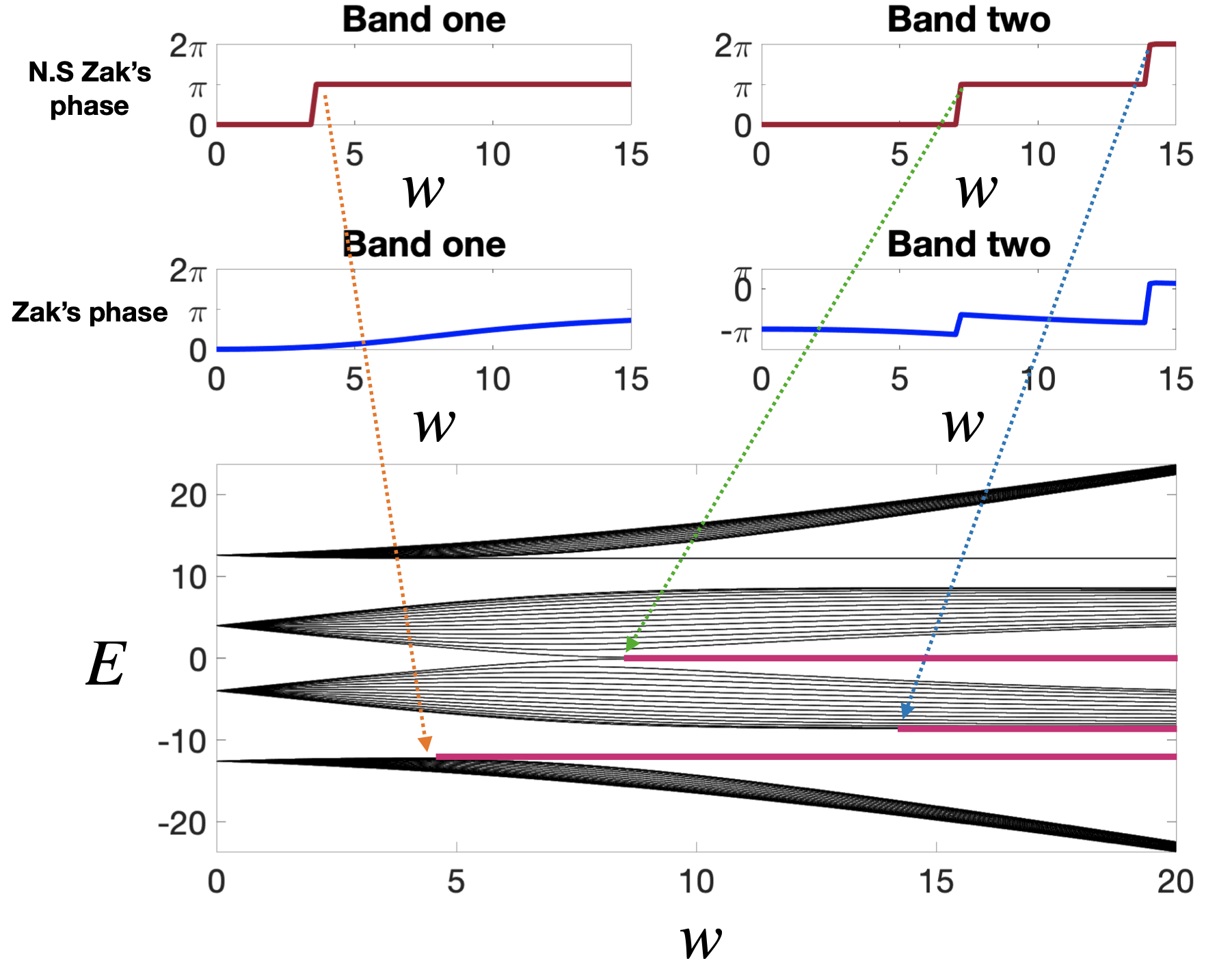}
            \caption{Comparison of NS Zak's phase with Zak's phase in the case of an SSH4 without mirror symmetry. While Zak's phase is not quantized, NS Zak's is quantized and predicts correctly the number of edge states that each band contributes. The spectral diagram corresponds to a finite chain with $4N$ (where $N = 20$) sites. Due to that, the NS Zak's corresponding to the first sublattice has been utilized. The bulk invariants for the first and second band are presented. The results for the rest of the bands are identical due to chiral symmetry.}
        \label{fig:sshfour_cont}
\end{figure}

\subsection{The case of $3N+2$ sites}

We turn now to the final case of $3N+2$ sites. In the chosen gauge, it is easy to observe that the phase of the sublattice $C$ is always zero. Nevertheless, the bulk-edge correspondence we have established also works for this case. That is, the corresponding (gauge-invariant) quantity $Z_{C,C}^\lambda - Z_{C,C}^{\lambda,\mathrm{ref}}$, which trivially vanishes, should also here be interpreted as the \textit{difference} in the number of edge states between the corresponding Hamiltonians. Indeed, for the case of $3N+2$ sites, the open chain always exhibits a constant number of edge states (i.e., two of the eigenstates are always edge states~\cite{he2018topology}), a fact that is reflected consistently in the vanishing of $Z_{C,C}^\lambda - Z_{C,C}^{\lambda,\mathrm{ref}}$. Clearly, the (nonvanishing) actual number of edge states present needs to be specified via a separate analysis, as was the case for $3N$ and $3N+1$ sites.

\section{Other SSH$m$ models}\label{sec::Other_sshm}

Last but not least, we point out that the introduced bulk quantity along with our interpretation of bulk-edge correspondence, works for the SSH model as well. Furthermore, NS Zak's phase turns out to be useful also for the SSH4 model, where it can be utilized to establish a bulk-edge correspondence even when mirror symmetry is absent.

In SSH (two sublattices) one can have two realizations of finite chains. One with even and one with odd number of sites (2N and 2N+1). In the gauge where: 
\begin{align}\label{ssh}
\ket{u_{\lambda}(k)}_{\mathrm{SSH}} = \frac{1}{\sqrt{2}}\begin{pmatrix}
a_{\lambda}^{A}(k)e^{-i\theta_{\lambda}^{A}(k)}\\
a_{\lambda}^{B}(k)\end{pmatrix}
\end{align}
one can see that NS Zak's phase of the first sublattice is identical with Zak's phase for this model. In fact, we have demonstrated that the phase of the first sublattice can establish a bulk-edge correspondence only for the case of an integer number of unit cells. That is the reason that the usual Zak's phase gives a well-defined bulk-edge correspondence only for the case of a chain with $2N$ sites. For the case of a chain with $2N+1$ sites, one should use the second sublattice according to our treatment. However, the phase of this sublattice vanishes since there is always an edge state present \cite{sirker2014boundary}  and thus there is \textit{no difference} in the number of edge states for different regimes of the parameters of the Hamiltonian.


As a last demonstration of the power of our approach we present some results for an SSH4, with an integer number of unit cells, that does not possess mirror symmetry. Specifically, Figure~\ref{fig:sshfour_cont} strongly suggests that NS Zak's phase can establish a bulk-edge correspondence even in regimes where the usual Zak's phase does not take integer values. In the same figure one can verify that NS Zak's phase correctly predicts the emergence of edge states for an SSH4 chain that does not possesses mirror symmetry. This suggests that NS Zak's phase is more general than the ordinary Zak's phase because it gives the correct results both in the presence and in the absence of mirror symmetry. Furthermore, it seems easily extendable for the general SSHm model.

\section{Discussion and Outlook}

We have thoroughly investigated the SSH3 model. We showed that, even in the absence of chiral and mirror symmetry, the fact that this model possesses point chiral symmetry has important consequences on the spectrum of the bulk Hamiltonian and the profile of the states.  We showed that NS Zak's phase is a good bulk quantity that can trace the change of the number of edge states in this system and establish a bulk-edge correspondence even for chains with noninteger number of unit cells ($3N$, $3N+1$, $3N+2$ sites). The framework that we presented allowed us to calculate the finite size corrections and provided a concrete way to count edge states even when the system possesses a relatively small number of unit cells. The fact that this model possess point chirality hints towards an equivalence of odd and even SSH$m$ models. This means that the existence of quantized bulk quantities in both cases may hint towards deeper common topological features of these models. Lastly, the derivation of bulk-edge correspondence that we provided in a relatively thorough manner for the case of SSH3 implies that similar derivations can be achieved for SSH$m$ models with $m$ sublattices.
Specifically, one can define NS Zak's phases for each of the $m$ sublattices and establish a bulk-edge correspondence for all the finite realizations of the SSH$m$ model (any noninteger cut of the unit cells). We have verified this numerically for SSH4. The advantage of this technique relative to others is that: a) it is simple -- one can predict the number of edge states for SSH$m$ without having to solve $m$-degree characteristic polynomials, b) It provides bulk-edge correspondence for noninteger number of unit cells, c) It gives a concrete path to calculate finite-size corrections. We leave as a future endeavor the treatment for the general SSH$m$ model.

\begin{acknowledgments}

The current project was funded by the Greek State Scholarships Foundation (I.K.Y) as part of Nikolaos D.~Xrysovergis grant. We thank P.~Delplace and M.~Bello Gamboa for comments on the manuscript. Special thanks to Li-Yang Zheng and Malte R\"{o}ntgen for fruitful conversations and critical remarks. G.S. acknowledges support through the ERC Advanced Grant QUENOCOBA No.~742102 and by the DFG (German Research Foundation) under Germany’s Excellence Strategy --  EXC-2111 -- 390814868. 

\end{acknowledgments}

\bibliography{bibliography_new}

\newpage

\appendix
  
\widetext


\section{Generalised Chirality and Point Chirality in SSH3} \label{sec:app:gen_chirality_point_chirality}

\subsection{Generalised Chirality}\label{sec:app:gen_ch}

Generalized chirality comes as a generalization of chiral symmetry -- instead of an operator that anticommutes with the Hamiltonian, one can find an operator that obeys generalized anticommutation relations. Specifically for the case of SSH3, in analogy to the chiral operator, a unitary operator $\Gamma_{g}$ having the following properties can be defined: 
\begin{subequations}
\begin{align}
    \Gamma^{3}_{g} &= 1 \;, \\
    H_{0} + H_{1} + H_{2} &= 0 \,,
\end{align}
\end{subequations}
where, by denoting the initial bulk Hamiltonian as $H_{0}$, 
\begin{align}\label{eq:gen_two}
   H_{1} \coloneqq \Gamma_{g}H_{0}\Gamma^{-1}_{g}  \nonumber\\
   H_{2} \coloneqq \Gamma_{g}H_{1}\Gamma^{-1}_{g}
\end{align}
%
\noindent For the case of SSH3, and the bulk Hamiltonian of \eqref{bulk_Ham}, it is easy to verify that an operator that obeys these relations exists, and is
\begin{align}
\Gamma_{g} = \diag ( 1 ,\omega, \omega^2 ) \,,
\end{align}
\noindent where $\omega \coloneqq e^{2 \pi i /3}$.

Using generalized chirality, one can define projectors over the sublattices of the unit cell
\begin{subequations}
\begin{align}
P_{A} &= \frac{1}{3}\left(I + \Gamma_{g} + \Gamma_{g}^{2}\right) \\
P_{B} &= \frac{1}{3}\left(I + \omega^2 \Gamma_{g} +  \omega \Gamma_{g}^{2}\right) \\
P_{C} &= \frac{1}{3}\left(I +  \omega \Gamma_{g} +  \omega^2  \Gamma_{g}^{2}\right) \,.
\end{align}
\end{subequations}
The Hamiltonian can be decomposed in the following manner
\begin{align}
    H = P_{A}HP_{B} + P_{B}HP_{C} + P_{C}HP_{A} + h.c.
\end{align}
\noindent and  the following holds
\begin{align}\label{eq:zero}
    P_{A}HP_{A} = P_{B}HP_{B} = P_{C}HP_{C} = 0 \,.
\end{align}
\noindent Relations \eqref{eq:zero} reflect the fact that the probability of a transition from one site to another one within the same sublattice is zero.

\subsection{Point Chirality}\label{sec:app:point_chiral}

As was already mentioned in the main text, SSH3 can be viewed as a degenerate SSH6 \cite{he2020non}. This means that instead of using~\eqref{eq:bulk_ham} in order to describe the system, one could use
\begin{align}\label{eq:degenerate_hex}
  H(k) = \begin{pmatrix}
  0 & u & 0 & 0 & 0 & we^{-ik}\\
   u & 0 & v & 0 & 0 & 0\\
     0 & v & 0 & w & 0 & 0\\
      0 & 0 & w & 0 & u & 0\\
        0 & 0 & 0 & u & 0 & v\\
         we^{ik} & 0 & 0 & 0 & v & 0\\
  \end{pmatrix}
\end{align}
which possesses chirality, i.e., a unitary and hermitian operator $\Gamma$ exists with $\Gamma H(k) \Gamma^{-1} = -H(k)$. The bulk spectrum of this Hamiltonian is essentially the zone folded spectrum of the SSH3 Hamiltonian. 


The degenerate points of SSH6 at the edges of the folded Brillouin zone correspond to the same points for SSH3 but for $k = \frac{\pi}{2}$. This means that these points will inherit some properties from SSH6 when one unfolds the spectrum in order to go to the SSH3. More specifically, since SSH6 posses chiral symmetry, SSH3 will have a point chiral symmetry at $k = \frac{\pi}{2}$. This was formulated in Eq.~\eqref{eq:point_symm}. We proceed now with the derivation of the consequences of point chirality for SSH3, which is exactly analogous to the case of ordinary chiral symmetry.





%

\begin{itemize}
\item \textit{Eigenstates always come in pairs with opposite energies.}
\end{itemize}

Assume $\ket{u(k)}$ is an eigenvector of $H_{\mathrm{bulk}}(k)$ with nonvanishing energy. Then,
\begin{align}\label{eq:app:proof_point_chirality}
    H_{\mathrm{bulk}}(k) \ket{u(k)} &= E(k) \ket{u(k)}\nonumber\\
     \Rightarrow \Gamma_{p}  H_{\mathrm{bulk}}(k)\Gamma_{p}^{\dagger} \Gamma_{p}\ket{u(k)} &= \Gamma_{p} E(k) \ket{u(k)}\nonumber\\
          \Rightarrow  H_{\mathrm{bulk}}(\pi+k) (\Gamma_{p}\ket{u(k)}) &= - E(k) (\Gamma_{p}  \ket{u(k)})
\end{align}
which means that $\Gamma_{p}  \ket{u(k)}$ is eigenvector of $H_{\mathrm{bulk}}(\pi+k)$ with eigenvalue $-E(k)$. The above derivation is also inherited to i) finite periodic chains with even sites (since $k+ \pi$ is an allowed wavenumber and hence point chirality is an exact symmetry in that case), and ii) finite open chains, since the latter can be considered as embedded within an infinite chain with appropriate boundary conditions.

\begin{itemize}
\item \textit{The partner eigenstates can be obtained from one another by the action of $\tilde \Gamma$.}
\end{itemize}

It follows directly from Eq.~\eqref{eq:app:point_chiral_extension_ssh3}.

\begin{itemize}
\item \textit{Eigenstates have equal support on even and odd sites.}
\end{itemize}

As is obvious from Eq.~\eqref{eq:app:point_chiral_extension_ssh3}, for the extended chain, point chirality takes the form of the familiar chiral operator. This means that two sub-lattice operators can be defined in the following manner: 
\begin{subequations} \label{chiral_sublattices}
\begin{align}
    P_{\mathrm{odd}} &= \frac{1}{2}(I +\tilde \Gamma)  \\
    P_{\mathrm{even}} &= \frac{1}{2}(I- \tilde \Gamma) \,,
\end{align}
\end{subequations}
and $\tilde \Gamma = P_{\mathrm{odd}} - P_{\mathrm{even}}$. From the orthogonality of states of the finite Hamiltonian follows that
\begin{align}
    \braket{\Psi|\tilde \Gamma_{p}|\Psi} = 0 \Rightarrow \braket{\Psi|P_{\mathrm{odd}}|\Psi} = \braket{\Psi|P_{\mathrm{even}}|\Psi}.
\end{align}
%

\section{Derivation of the momentum shift equations} \label{sec:app:momentum_shift}

Here we show in detail how one can derive Eq.~\eqref{central} following closely the approach of Ref.~\cite{marques2020analytical}. Our starting point is the Hamiltonian
\begin{align}\label{Hamiltonian}
H =- \begin{pmatrix}
0 & v & 0 & 0 & 0 & \dots\\
v & 0 & u & 0 & 0 & \dots\\
0 & u & 0 & w & 0 & \dots\\
\vdots& \vdots & \ddots & \ddots& \ddots & \dots\\
0 & \dots & 0 & 0 & 0 & 0 &w & 0 & 0  \\
0 & \dots & 0 & 0 & 0 & w &0 & v & 0 \\
0 & \dots & 0 & 0 & 0 & 0&v & 0 & u \\
 0 & \dots & 0 & 0 & 0 & 0 &0 & u& 0  \\
\end{pmatrix}
\end{align}
for the case of SSH3. Our derivation is for a system with open boundary conditions and an integer number of unit cells. For later convenience, the (ordered) basis upon which \eqref{Hamiltonian} is written is $\left( \ket{3N}, \ket{3N-1},\dots \ket{2},\ket{1} \right)$ with $\ket{j}$ being the $j$th site of the chain. Our aim is to solve the eigenvalue-eigenvector problem for this Hamiltonian. The characteristic polynomial of this problem is denoted as
\begin{align}
x_{1:3N}(E_\lambda) = \mathrm{det}(E_{\lambda} I-H) \;,
\end{align}
\noindent where we use a bottom up notation for the determinant (i.e., $x_{1:1}$ is the determinant of a chain with a single particle at position $3N$ and $x_{1:3N}$ is the determinant of a complete chain with all $3N$ sites). From the form of the Hamiltonian~\eqref{Hamiltonian}, it follows that we can expand
\begin{align}\label{re_char_pol}
x_{1:3N}(E_{\lambda}) = E_{\lambda} x_{1:3N-1} - u^{2}x_{1:3N-2} \,.
\end{align}
We observe that if we would like to write a corresponding relation for $x_{1:3N-1}$, $v$ would appear on the r.h.s instead of $u$. For $x_{1:3N-2}$, $w$ would appear instead of $u$ or $v$ and for $x_{1:3N-3}$ we would return to the initial relation. This motivates us to introduce the following notation: We define $x_{n}^{i}(E_{\lambda}) \coloneqq x_{1:3n+1-i}(E_{\lambda})  $, where $i = 1,2,3$ and $n = 0,1,2, \dots,N$ where $x_{0}^{i}$ is to be determined by the boundary conditions.

By using this notation, we can rewrite (\ref{re_char_pol}) as a set of coupled equations: 
\begin{align}\label{system}
x_{n}^{1}(E_{\lambda})  &= E_{\lambda}x_{n}^{2} - u^{2} x_{n}^{3}\nonumber\\
x_{n}^{2}(E_{\lambda})  &= E_{\lambda} x_{n}^{3} - v^{2} x_{n-1}^{1}\nonumber\\
x_{n}^{3}(E_{\lambda})  &= E_{\lambda} x_{n-1}^{1} - w^{2} x_{n-2}^{2}
\end{align}

By doing a little bit of algebra, we arrive at
\begin{align}\label{firstsublattice}
x_{n}^{1}(E_{\lambda})  =   \left(E_{\lambda}^{3} - E_{\lambda}(u^{2}+v^{2}+w^{2})\right)x_{n-1}^{1}(E_{\lambda}) - (uvw)^{2}x_{n-2}^{1}(E_{\lambda})  \,.
\end{align}

The next step is to utilize the bulk solutions for $E_{\lambda}(k)$ and use this as a parameter in (\ref{firstsublattice}). The bulk Hamiltonian reads
\begin{align}\label{bulk_Ham}
H_{\mathrm{bulk}}(k) = \begin{pmatrix}
0 & u & we^{-ik}\\
u & 0 & v \\
w^{ik} & v & 0\\
\end{pmatrix}
\end{align}
yielding the eigenvalue problem
\begin{align}\label{char_bulk}
\mathrm{det}\left(E_{\lambda}I - H_{\mathrm{bulk}}(k)\right) &= 0  \nonumber\\
E_{\lambda}^{3} - E_{\lambda}(u^{2}+v^{2}+w^{2}) + 2uvw \cos(k) &= 0 \,.
\end{align}
By comparing (\ref{char_bulk}) and (\ref{firstsublattice}) we see that we can get
\begin{align}\label{bulk_rec_rel}
x_{n}^{1}(E_{\lambda}) = -2uvw\cos(k)x_{n-1}^{1} - (uvw)^{2}x_{n-2}^{1}
\end{align}
from which, all $x_{n\geq 2}^{1}(E_{\lambda})$ can be obtained if $x_{0}^{1}$ and $x_{1}^{1}(E_{\lambda})$ are known. We set as a boundary condition $x_{0}^{1}(E_{\lambda}) = 1$ and thus
\begin{align}\label{border}
x_{1}^{1}(E_{\lambda}) &= x_{1:3}(E_{\lambda}) = \mathrm{det}\begin{pmatrix}
E_{\lambda} & u & 0 \\
u &E_{\lambda} & v \\
0 & v & E_{\lambda}\\
\end{pmatrix} \nonumber \\
&= E_{\lambda}(E_{\lambda}^{2}-v^{2}) -u^{2}E_{\lambda}= E_{\lambda}^{3} - E_{\lambda}(u^{2}+v^{2})
\end{align}
where we can use again (\ref{char_bulk}) and arrive at
\begin{align}
x_{1}^{1}(\lambda) = -2uvw\cos(k) +E_{\lambda} w^{2} \,.
\end{align}
If we set $T = -uvw$, then (\ref{bulk_rec_rel}) takes the form 
\begin{align}\label{firstsub}
x_{n}^{1}(E_{\lambda}) = 2T\cos(k) x_{n-1}^{1}(E_{\lambda}) - T^{2} x_{n-2}^{1}(E_{\lambda})
\end{align}
and if we redefine $W_{n}^{1}(E_{\lambda},\cos(k)) = T^{N-n}x_{n}^{1}(E_{\lambda})$, then (\ref{firstsub}) becomes
\begin{align}\label{secondsub}
W_{n}^{1} = 2\cos(k)W_{n-1}^{1} - W_{n-2}^{1}
\end{align}
which is the recurrence relation of the Chebyshev polynomials of second kind. Specifically the Chebyshev polynomials of the second kind are defined in the following manner:
\begin{align}\label{Chebysev}
U_{0}(x) &= 1\\
U_{1}(x) &= 2x\\
U_{n+1} &= 2xU_{n}(x) - U_{n-1}(x)
\end{align}

This is the relevant case for our purposes with $U_{n} = U_{n}(\cos(k))$ and an extra boundary condition $U_{-1} = 0$. From the definition of $W_{n}^{1}(E_{\lambda},\cos(k))$ we have
\begin{align}
W_{0}^{1} &= T^{N}(1 +0) = T^{N}(U_{0} + U_{-1})\\
W_{1}^{1} &= T^{N}(2\cos(k) +a^{1}_{\lambda}(k)) = T^{N}(U_{1} +a^{1}_{\lambda}(k)U_{0})
\end{align}
where $a^{1}_{\lambda}(k) = -\frac{w}{uv}E_{\lambda}(k)$. So we arrive at the recursive relation:
\begin{align}\label{finalsub}
W_{n}^{1} = T^{N}(U_{n}(\cos(k)) + a^{1}_{\lambda}(k) U_{n-1}(\cos(k))) \,.
\end{align}

Now, our initial aim was to solve $x_{N}^{1}(E_{\lambda}) = 0$ which translates to $W_{N}^{1}(E_{\lambda},\cos(k)) = 0$. Using (\ref{finalsub}) and the known identity $U_{n}(\cos(k)) = \frac{\sin[(n+1)k]}{\sin(k)}$ one gets
\begin{align}
\frac{\sin[(N+1)k]}{\sin(k)} + a^{1}_{\lambda}(k)\frac{\sin(Nk)}{\sin(k)} = 0
\end{align}
which, with some trigonometry transforms to
\begin{align}
\cot[(N+1)k] = \frac{1}{a^{1}_{\lambda}(k)\sin(k)} +\cot(k)
\end{align}

As a result, we arrive at condition (\ref{central}). One may find the general way to extract the corresponding relations for the general SSH$m$ model and also for the case with non-integer unit cells in Ref.~\cite{marques2020analytical}.

\section{Analytical expressions for $\theta^A_\lambda (k)$, $\theta^B_\lambda (k)$ and the equation $g_\lambda(k) = 0$} \label{sec:app:trigonometric}

Here, we first derive an explicit expression for $\theta^A_\lambda (k)$ and $\theta^B_\lambda (k)$ as a function of the hopping parameters. The characteristic polynomial of the bulk Hamiltonian reads
\begin{align}\label{char_pol}
E_{\lambda}^{3} - E_{\lambda}(u^{2}+v^{2}+w^{2}) + 2uvw \cos(k) =  0\,.
\end{align}
\noindent This is the form of a depressed cubic, i.e., a polynomial of the third degree in the form
\begin{align}\label{depressed_cubic}
 t^{3}+pt+q=0 \,.
\end{align}

\noindent If all three roots are real (which are in the case we are interested in since our Hamiltonian is hermitian) the solutions can be written in the trigonometric form
\begin{align}\label{depressed_cubic_sol}
 t_{k}=2\,{\sqrt {-{\frac {p}{3}}}}\,\cos \left[\,{\frac {1}{3}}\arccos \left({\frac {3q}{2p}}{\sqrt {\frac {-3}{p}}}\,\right)-{\frac {2\pi k}{3}}\,\right]\qquad {\text{for }}k=0,1,2.
\end{align}

\noindent By comparing \eqref{char_pol} and \eqref{depressed_cubic} we see that $p = -(u^{2}+v^{2}+w^{2})$ and $q = 2uvw \cos(k)$. Thus the expression for $a^{1}_{\lambda} = -\frac{w}{uv} E_{\lambda}(k)$ is given by
\begin{align}
    a^1_\lambda = -\sqrt{\frac{4}{3}} \frac{\sqrt{1+a^2+b^2}}{a b}\cos \left[ \frac{1}{3} \arccos \left(  - \frac{a b \cos k}{\left(  \frac{a^2+b^2+1}{3}\right)^{3/2}} \right) - \frac{2 \pi \lambda}{3}  \right] \,,
\end{align}

\noindent where $\lambda = 0,1,2$ enumerates the bands from bottom to top and also we have expressed $a:=\frac{u}{w}$, $b:=\frac{v}{w}$. As a result, is it easy to check that the following choice makes the angles $\theta_\lambda^A (k)$ continuous (and differentiable) in $(0,\pi)$:
\begin{subequations} \label{eq:theta_full_form}
\begin{align}
    \theta^A_{\lambda} (k) = \arccot \left( \frac{1}{a_\lambda^1 \sin k} + \cot k \right) \,, \qquad \lambda = 0,2
    \end{align}
and
    \begin{align}
    \theta^A_{\lambda} (k) = \begin{cases} 
          \arccot \left( \dfrac{1}{a_\lambda^1 \sin k} + \cot k \right) & k \in (0, \pi/2 ]  \\
          \\ \arccot \left( \dfrac{1}{a_\lambda^1 \sin k} + \cot k \right) + \pi & k \in ( \pi/2 , \pi ]
       \end{cases} \;,
       \qquad \lambda = 1
\end{align}

\end{subequations}

\noindent where $\arccot$ takes values in $(0, \pi)$.

Let us now explain why the equation $g_\lambda(k) = 0$ determines all hyperspaces which separate regions with (possibly) different values of $Z_{A,C}^\lambda$.

First of all recall that, due to the divergence of Eq.~\eqref{central_2} for $k \to 0^+,\pi^-$, we concluded that $\theta_\lambda^A(k)$ is an integer multiple of $\pi$ at these two points. Moreover, the sign of the divergence of Eq.~\eqref{central_2} as $k \to 0^+$ has to agree with the sign of the derivative $d\theta_\lambda^A(k) /d k$ as $k \to 0$, while the two signs have to be opposite at $k \to \pi$. In turn, from Eqs.~\eqref{eq:theta_full_form} above one can read that, for each band, there are at most two different values for $\theta_\lambda^A(k)$ at $k = 0$ and different values always come with a different sign in the derivative. A similar observation holds for $k = \pi$.

Combing these two last facts, we reach the conclusion that \textit{a change in the value of the NS Zak's phase $Z_{A,C}^\lambda$ is necessarily associated with a change in the sign of the divergence of Eq.~\eqref{central_2}, either at $k \to 0^+$ or $k \to \pi^-$}. In turn, this sign is determined by the numerator $g_{\lambda}(k) \coloneqq 1 - \frac{w}{u v} E_{\lambda}(k) \cos(k)$. This is because the energy $E_\lambda(k)$ of each band has a fixed sign at $k=0, \pi$. Finally, due to continuity, a sign changes can thus only occur when $g_{\lambda}(k)=0$.

For the case of $3N+1$ sites, instead of $a_{1}^\lambda$ one needs the expression of $a_{2}^\lambda$. The corresponding formula will be given by $a_{2}^\lambda = -\frac{uw}{v}\frac{1}{E_{\lambda}(k)}$ and one can extract it by following the method exhibited in Appendix~\ref{sec:app:momentum_shift} but by adding an extra site to the initial Hamiltonian. For a more detailed treatment of the momentum shift for non integer number of unit cells, see~\cite{marques2020analytical}. Then, in order to get $\theta^B_{\lambda} (k)$, one uses the relations given in \eqref{eq:theta_full_form} but with the substitution of $a_{1}^\lambda$ with $a_{2}^\lambda$.

\end{document}